\documentclass[%
 reprint,
 amsmath,amssymb,
 aps,prx,
]{revtex4-2}

\usepackage{dcolumn}% Align table columns on decimal point
\usepackage{bm}% bold math
\usepackage{hyperref, soul}
\hypersetup{linktocpage,,colorlinks=true}
\usepackage{hyperref, soul}
\usepackage{physics}
\usepackage{subcaption}
\usepackage{graphicx}
\usepackage{cleveref}
\usepackage{tikz}
\usepackage{verbatim}
\usepackage{caption}
\makeatletter
\newcommand{\SM@oldaddcontentsline}{} % storage

\newcommand{\SMtocredirectON}{%
  \let\SM@oldaddcontentsline\addcontentsline
  \renewcommand{\addcontentsline}[3]{%
    \def\SM@file{##1}%
    \def\SM@toc{toc}%
    \ifx\SM@file\SM@toc
      \SM@oldaddcontentsline{smtoc}{##2}{##3}%
    \else
      \SM@oldaddcontentsline{##1}{##2}{##3}%
    \fi
  }%
}

\newcommand{\SMtocredirectOFF}{%
  \let\addcontentsline\SM@oldaddcontentsline
}

\newcommand{\SMtableofcontents}{%
  \begingroup
    \setcounter{tocdepth}{2}% 1=section,2=subsection
    \par\noindent{\bfseries Contents}\par\medskip
    \@starttoc{smtoc}%
  \endgroup
}

\newcommand{\panel}[3]{%
  \begin{tikzpicture}[inner sep=0pt, outer sep=0pt]
    \node[anchor=south west, inner sep=1.5pt] (img) at (0,0)
      {\includegraphics[width=\dimexpr #2-3pt\relax]{#1}}; % 2pt left + 2pt right
    \node[anchor=north west, font=\bfseries, fill=white, rounded corners=0pt, inner sep=0pt]
      at ([xshift=0pt,yshift=-0.1pt]img.north west) {#3};
  \end{tikzpicture}%
}

\makeatother
\begin{document}

\preprint{APS/123-QED}

\title{%Toward a dS/(c)MERA Correspondence from Non-Hermitian Critical Systems
Emergent de Sitter Space and Non-Unitary Tensor Networks from Non-Hermitian Quantum Criticality
}% Force line breaks with \\

\author{Kuang-Hung Chou} \email{nagisa.7256.960247@gmail.com}
\affiliation{%
 Department of Physics, National Tsing Hua University, Hsinchu 30013, Taiwan
}
\author{Po-Yao Chang}%
 \email{prayser@gmail.com}
\affiliation{%
 Department of Physics, National Tsing Hua University, Hsinchu 30013, Taiwan
}%

\date{\today}% It is always \today, today,

\begin{abstract}

Extending the holographic principle to de Sitter (dS) spacetimes remains one of the most vital open frontiers in quantum gravity, where a microscopic, bottom-up tensor-network framework that relates boundary quantum data to emergent de Sitter spacetime is still lacking. In this work, we first show the emergence of de Sitter spacetime from boundary entanglement by formulating a non-unitary continuous multi-scale entanglement renormalization ansatz (cMERA) for a concrete non-Hermitian critical fermion chain. Within this emergent spacetime, we analyze the associated geodesics and show that they act as extremal Ryu-Takayanagi (RT) surfaces undergoing a smooth timelike-to-null transition. Remarkably, we demonstrate that this continuum trajectory suggests a distinct tensor-network architecture in which the bond-counting contribution naturally truncates at the discrete timelike-to-null transition toward the deep infrared. In the resulting architecture, the null ray along the horizon is represented by zero-cost links, since the associated cut severs no tensor legs. This network structure successfully reproduces the logarithmic scaling of non-unitary critical entanglement entropy, offering a bond-counting picture for the de Sitter RT formula. Our results provide the long-sought dS/(c)MERA correspondence at the level of both emergent spacetime and discrete holographic entanglement.
\end{abstract}

\maketitle

%\tableofcontents

\section{Introduction}
\label{sec:intro}

Tensor-network architectures have emerged as a powerful framework for deciphering the structural mechanics of quantum gravity, offering microscopic models for aspects of the anti-de Sitter/conformal field theory (AdS/CFT) correspondence~\cite{Maldacena99,GUBSER1998105,Edward1998}. On the geometric side, the Ryu-Takayanagi (RT) formula maps the entanglement entropy $S_A$ of a boundary subregion $A$ to the area of a bulk minimal surface $\gamma_A$~\cite{PhysRevLett.96.181602,Hubeny2007HRT,Lewkowycz2013GeneralizedEntropy} 
\begin{equation}
S_A=\frac{\mathrm{Area}(\gamma_A)}{4G_N}
\end{equation}
Conversely, on the quantum information side, the multi-scale entanglement renormalization ansatz (MERA) implements real-space entanglement renormalization~\cite{PhysRevLett.99.220405,PhysRevLett.101.110501,PhysRevB.79.144108} where the entropic cost of a boundary subregion is upper-bounded by a discrete minimal cut through the network layer connectivity
\begin{equation}
S_A \lesssim \min_{\mathcal C_A} n(\mathcal C_A)\,\log \chi ,
\label{eq:mincut}
\end{equation}
where \(n(\mathcal C_A)\) is the number of bonds cut by the curve \(\mathcal C_A\) and \(\chi\) is the bond dimension. This discrete bond-counting prescription yields an elegant graph-theoretic mirror to the continuum RT formula~\cite{AdS_MERA,AdS_MERA_2}, providing the foundation for the AdS/MERA correspondence~\cite{Evenbly2011TensorNetworkGeometry,Milsted:2018san,PhysRevX.15.021078}. In the continuum limit, the continuous MERA (cMERA) formalism~\cite{Haegeman2013} demonstrates how a spatial radial direction dynamically emerges from the scale-dependent entangling structure of a boundary quantum state, transforming the renormalization group (RG) flow into the warped spatial slices of Poincar\'{e} AdS~\cite{Nozaki2012}. More broadly, exact holographic mappings, random tensor networks, and holographic-code tensor networks, together with the quantum-error-correcting-code interpretation of holography, have sharpened how geometric entanglement rules and bulk encoding can arise from microscopic boundary structures~\cite{Pastawski2015,Hayden2016,Almheiri2015,Harlow2017,qi2013exactholographicmappingemergent,Yang2016BidirectionalHolographicCodes,Steinberg2023HyperinvariantHolographicCodes,Dong2024,PhysRevD.105.026018}.

Extending this holographic dictionary to de Sitter (dS) spacetimes remains one of the most vital yet conceptually elusive frontiers in quantum gravity. In de Sitter holography (dS/CFT), the fundamental nature of boundary unitarity, spatial conjugation, and asymptotic observables diverges qualitatively from the AdS paradigm~\cite{witten2001quantumgravitysitterspace,Strominger2001,Anninos2017,Hikida2022,Doi2025}. While innovative top-down tensor networks~\cite{PhysRevD.96.123536,PhysRevD.105.125009,Cao2025} have been proposed to capture the kinematic structure and horizon organization of de Sitter space, a direct bottom-up counterpart to the AdS/(c)MERA correspondence remains less developed. Specifically, the field lacks a microscopic formulation where a concrete boundary many-body system explicitly generates both an emergent de Sitter geometry and a discrete bond-counting framework that faithfully reflects the RT surface profile.

Recent developments at the boundary of quantum information and non-Hermitian physics suggest a natural resolution to this challenge. Recent  analyses indicate that de Sitter geometries can be extracted from boundary CFT state data via the information metric, provided the framework adopts a future-past conjugation structure that structurally mirrors the left-right (biorthogonal) state pairing native to non-Hermitian or non-unitary systems~\cite{Doi2025}; related non-Hermitian density-matrix structures have also been discussed in recent de Sitter holography contexts~\cite{PhysRevLett.130.031601,harper2026nonhermitiandensitymatricestimelike,Anastasiou2026RenormalizedPseudoentropy}. Concurrently, non-Hermitian critical chains governed by biorthogonal density matrices have provided concrete, microscopic lattice realizations of non-unitary quantum criticality~\cite{nH_SSH,Renyi,chou2026ptsymmetryenrichednonunitarycriticality,zhou2024universalnonhermitianflowonedimensional,io2026nonhermitianfreefermioncriticalsystems,Shimizu:2025kse,PhysRevB.107.205153,rottoli2024entanglementhamiltoniannonhermitianssh}.

In this work, we systematically realize this program by constructing a bottom-up dS/(c)MERA framework from a concrete non-Hermitian critical fermion chain. Our construction has three main components:

\begin{itemize}

\item {\bf Microscopic Derivation of Global de Sitter Space:} By formulating a non-unitary cMERA for a critical non-Hermitian Su-Schrieffer-Heeger model~\cite{Klett2017, Lieu2018}, we show that the information metric associated with the continuous entanglement-renormalization flow acquires Lorentzian signature~\cite{Fubini}. Crucially, the RG scale direction behaves as a timelike coordinate, yielding an emergent \((1+1)\)-dimensional de Sitter spacetime directly from boundary critical data.

\item {\bf Natural Embedding into the Penrose Diagram:} This bottom-up construction naturally maps onto the global causal structure of the dS Penrose diagram. The past and future conformal boundaries ($\mathcal{I}^\pm$) correspond to the left and right boundary UV states of the two-ended cMERA circuit. As these coupled boundary states flow toward the deep infrared, they converge onto a unique, single-site IR endpoint characterized by a non-entangled non-Hermitian reference density matrix.

\item {\bf RT-Surface Interpretation via Alternative Bond Counting:} Within this emergent geometry, we demonstrate that the relevant de Sitter geodesic acts as a generalized RT surface, connecting the boundary interval endpoint to the deep-IR pole. We show that this continuum geodesic undergoes a smooth timelike-to-null transition that can be mapped onto a discrete tensor-network skeleton via a modified bond-counting dictionary.
\end{itemize}

Crucially, our discrete dS/MERA skeleton suggests that the network structure needs to be intrinsically interval-centered. To correctly replicate the entropic profile of the RT surface, the site placement naturally excludes the interior of the observer-dependent static patch, leaving it entirely devoid of physical tensor-network degrees of freedom. The null ray along the static-patch horizon is instead represented by zero-cost links that sever no tensor legs.

This interval-dependent architecture successfully reproduces the expected logarithmic scaling of non-unitary critical entanglement entropy \((S_A \propto i\ln l_A)\), and suggests an interval- or observer-dependent error-correcting structure for de Sitter tensor networks, in contrast to the fixed bulk-encoding structure familiar from AdS tensor-network models.

The remainder of this paper is organized as follows. In Sec.~\ref{sec:ER}, we review the conventional cMERA construction and its established AdS benchmark. In Sec.~\ref{sec:NHcMERA}, we introduce the non-Hermitian free fermionic lattice chain and derive its emergent de Sitter metric from the non-unitary cMERA flow. Sec.~\ref{sec:RT} evaluates the continuum Lorentzian geodesics and establishes the geometric lessons governing de Sitter extremal surfaces. In Sec.~\ref{sec:discrete_MERA}, we translate these rules into a discrete graph-metric language, constructing the tensor network and validating its bond-counting interpretation. We present concluding remarks and future outlooks in Sec.~\ref{sec:con}.

\section{Continuous multi-scale entanglement renormalization ansatz and emergent AdS geometry}
\label{sec:ER}
We begin by reviewing cMERA for Hermitian critical fermions and the resulting AdS/cMERA interpretation.
The non-Hermitian generalization is developed in Sec.~\ref{sec:NHcMERA}; readers familiar with cMERA may proceed directly there.

The cMERA framework formalizes state preparation via a scale-ordered entangling flow driven along a continuous RG trajectory $u$, serving as the continuum analogue of the discrete MERA structure.
Compared with the discrete MERA description, where alternating discrete layers of disentanglers and coarse graining are required, the cMERA implements an infinitesimal RG step generated by (i) a scale transformation and (ii) a disentangling operation that injects correlations appropriate to that scale. We adopt the standard convention that the UV state sits at $u=0$, while the deep IR corresponds to $u\to -\infty$. The flow starts from an entangled UV state and disentangles down to an unentangled IR reference.
\begin{equation}
\ket{\Psi(u=-\infty)}=\ket{\Omega}, \quad \ket{\Psi(u=0)}=\ket{\Psi_{\mathrm{UV}}}.
\label{eq:cmera_bdy}
\end{equation}
The evolution is implemented by the path-ordered unitary
\begin{equation}
\ket{\Psi(u)}
=
\mathcal{\tilde{P}}\exp\!\left(-i\int_{u}^{0}\big[K(s)+L\big]\,ds\right)\ket{\Psi_{\mathrm{UV}}},
\label{eq:cmera_flow}
\end{equation}
where $L$ generates scale transformations and $K(u)$ is the disentangler. The scale generator is taken in the standard continuum form
\begin{equation}
L
=
-\frac{i}{2}\int dr\,
\Big(\psi^{\dagger}(r)\,r\,\partial_{r}\psi(r)
-
r\,\partial_{r}\psi^{\dagger}(r)\,\psi(r)\Big),
\label{eq:cmera_L}
\end{equation}
so that it implements the expected dilation on fields in position and momentum space,
\begin{align}
e^{-iuL}\psi(r)e^{iuL}
&=
e^{u/2}\psi(e^{u}r),\label{eq:cmera_dilate}\\
e^{-iuL}\psi(k)e^{iuL}
&=
e^{-u/2}\psi(e^{-u}k).
\end{align}
For noninteracting (bilinear) settings we record the canonical disentangler used throughout cMERA constructions,
\begin{equation}
K(u)
=
i\int dk\,
\Big(g_k(u)\,\psi^{\dagger}_1(k)\psi_2(k)+\mathrm{h.c.}\Big),
\label{eq:cmera_K_bilinear}
\end{equation}
where the scale profile $g_k(u)$ is fixed variationally or by matching correlation functions.

To implement the coarse-graining aspect of entanglement renormalization in the continuous setting, it is standard to impose a UV momentum cutoff and restrict the disentangler to act only on modes within this window. Operationally, this is encoded by multiplying the disentangler profile by a cutoff function $\Gamma$. A convenient choice is the sharp cutoff
\begin{equation}
\Gamma\!\left(\frac{|k|}{\Lambda}\right)=\Theta\!\left(1-\frac{|k|}{\Lambda}\right),
\label{eq:cmera_Gamma}
\end{equation}
so that the disentangler acts only on modes satisfying
\begin{equation}
|k|<\Lambda.
\label{eq:cmera_active_window}
\end{equation}
For a lattice model, the cutoff momentum is usually chosen as $\Lambda=\pi/a$, where $a$ is the lattice spacing. Accordingly, the coefficient in Eq.~\eqref{eq:cmera_K_bilinear} is understood to include the cutoff factor, $g_k(u)\rightarrow g_k(u)\,\Gamma\!\left(\frac{|k|}{\Lambda}\right)$.
In this way, as one flows toward the IR ($u\to-\infty$), the set of modes that are rotated by the disentangler shrinks continuously. Modes outside the cutoff window are not acted on by the disentangler. 
Accordingly, when we later quantify the local entangling strength using the \emph{interaction-picture} overlap between neighboring layers, only the active sector contributes nontrivially; the eliminated (coarse-grained) degrees of freedom play no role, in direct analogy with the discarded ancillas in discrete MERA.

For later calculations it is convenient to work in the interaction picture with respect to the scale generator $L$. Define the interaction-picture state and disentangler by
\begin{align}
\ket{\Psi^{I}(u)}&\equiv e^{iuL}\ket{\Psi(u)},\\
\widetilde{K}(u)&\equiv e^{iuL}K(u)e^{-iuL}.
\label{eq:cmera_int}
\end{align}
In this picture the state evolves only by the transformed disentangler,
\begin{equation}
\ket{\Psi^{I}(u)}
=
\mathcal{\tilde{P}}\exp\!\left(-i\int_{u}^{0}\widetilde{K}(s)\,ds\right)\ket{\Psi_{\mathrm{UV}}}.
\label{eq:cmera_int_flow}
\end{equation}
In the remainder of this work, unless otherwise stated, we will work in the interaction picture and simply write $\ket{\Psi(u)}$ for the interaction-picture state $\ket{\Psi^{I}(u)}$.

In direct analogy with the MERA picture---where bulk distance is tied to how many entangling bonds are cut and the dimension of each cut bond---the cMERA flow provides a continuous measure of the effective ``bond dimension'' encountered at each RG step. Infinitesimally, the disentangler rotates the many-body state as $u\to u+du$, and this change is captured by the overlap between neighboring layers. Using the interaction-picture states defined above, we quantify the local strength of entanglement renormalization along the RG direction by the Fubini--Study line element,
\begin{equation}
g_{uu}(u)\,du^2
\equiv
1-\big|\braket{\Psi(u)}{\Psi(u+du)}\big|^{2}.
\label{eq:cmera_FS}
\end{equation}
Conceptually, a stronger entangling action produces a more rapid rotation of the state in Hilbert space, hence a smaller overlap $\big|\braket{\Psi(u)}{\Psi(u+du)}\big|$ and a larger $g_{uu}(u)$. In this way, the ``bond dimension'' extracted from the overlap plays the continuous counterpart of the MERA bond-counting intuition.

To complete the geometric interpretation, one must also specify how the boundary spatial direction is represented at different RG layers. In cMERA, changing $u$ corresponds to coarse graining, so physical lengths rescale under the dilation generated by $L$. Using Eq.~\eqref{eq:cmera_dilate}, a boundary coordinate separation transforms as $x\mapsto e^{u}x$, implying that the spatial line element at layer $u$ should acquire a factor $e^{2u}$. This leads to the standard warped form
\begin{equation}
ds^{2}
=
g_{uu}(u)\,du^{2}
+
\Lambda^{2}e^{2u}\,dx^{2},
\label{eq:cmera_line_element}
\end{equation}
with the UV scale $\Lambda$ fixing the units of the boundary coordinate $x$. For unitary critical systems, $g_{uu}(u)$ approaches a positive constant in the scale-invariant regime. Introducing the radial coordinate $z\propto e^{-u}$ then brings Eq.~\eqref{eq:cmera_line_element} into the Poincar\'e form of an AdS spatial slice,
\begin{equation}
ds^{2}
=
\frac{g_{uu}}{z^{2}}\big(dz^{2}+dx^{2}\big),
\label{eq:cmera_AdS}
\end{equation}
thereby recovering the AdS/cMERA correspondence as the continuous counterpart of the AdS/MERA picture.

\section{cMERA for non-Hermitian free fermionic systems}
\label{sec:NHcMERA}

\subsection{Model}
\label{subsec:model_short}

% --- Right/Left spinor macro definitions
\def\Lp{
	\begin{bmatrix}
		\cos^{\ast}\!\big(\tfrac{\phi_k}{2}\big)\\
		\sin^{\ast}\!\big(\tfrac{\phi_k}{2}\big)
\end{bmatrix}}
\def\Lm{
	\begin{bmatrix}
		-\sin^{\ast}\!\big(\tfrac{\phi_k}{2}\big)\\
		\cos^{\ast}\!\big(\tfrac{\phi_k}{2}\big)
\end{bmatrix}}
\def\Rp{
	\begin{bmatrix}
		\cos\!\big(\tfrac{\phi_k}{2}\big)\\
		\sin\!\big(\tfrac{\phi_k}{2}\big)
\end{bmatrix}}
\def\Rm{
	\begin{bmatrix}
		-\sin\!\big(\tfrac{\phi_k}{2}\big)\\
		\cos\!\big(\tfrac{\phi_k}{2}\big)
\end{bmatrix}}

We focus on the two-legged non-Hermitian SSH model defined in momentum space by
\begin{equation}
H=\sum_k
\begin{bmatrix}
\psi_1^\dagger(k) & \psi_2^\dagger(k)
\end{bmatrix}
\mathcal{H}_k
\begin{bmatrix}
\psi_1(k)\\
\psi_2(k)
\end{bmatrix}
\end{equation}
with
\begin{equation}
\mathcal{H}_k=
\begin{bmatrix}
i\mu & -w\cos(ka)+v\\
-w\cos(ka)+v & -i\mu
\end{bmatrix},
\end{equation}
where $k$ is the single-particle momentum, $a$ is the lattice spacing, and $\mu,v,w\in\mathbb{R}^+$. The model is critical when the gap closes at $k=0$, i.e., $|v-w|=\mu$; throughout we focus on the critical line $v-w=\mu$.

Since $\mathcal{H}_k$ is non-Hermitian, we have different eigenstates at the left and right,
\begin{equation}
\mathcal{H}_k\ket{R_\alpha}=E_{\alpha}\ket{R_\alpha},\quad
\bra{L_\alpha}\mathcal{H}_k=E_\alpha\bra{L_\alpha},
\end{equation}
with the biorthonormality condition $\braket{L_{\alpha}}{R_{\beta}}=\delta_{\alpha\beta}$. The right eigenstates are
\begin{equation}
\ket{R_{k,+}}=\Rp,\quad \ket{R_{k,-}}=\Rm,
\end{equation}
with complex angle
\begin{equation}
\phi_k=\arctan\!\left(\frac{-w\cos(ka)+v}{i\mu}\right),
\label{eq:phase}
\end{equation}
where $\phi_k$ is chosen on the principal branch. The corresponding left eigenstates are
\begin{equation}
\ket{L_{k,+}}=\Lp,\quad \ket{L_{k,-}}=\Lm.
\end{equation}
At half filling, the many-body ground states are Gaussian products,
\begin{equation}
\ket{\Psi_R}=\prod_k \psi_{R,-}^\dagger(k)\ket{0},
\quad
\bra{\Psi_L}=\bra{0}\prod_k \psi_{L,-}(k),
\end{equation}
where $\psi_{R,-}^\dagger(k)$ creates a particle in $\ket{R_{k,-}}$ and $\psi_{L,-}(k)$ annihilates a particle in the corresponding left mode, with $\braket{\Psi_L}{\Psi_R}=1$. We then define the biorthogonal density matrix
\begin{equation}
\rho=\ket{\Psi_R}\bra{\Psi_L}.
\end{equation}
Because $\rho$ is non-Hermitian, entanglement entropies defined from $\rho$ need not be positive definite; in particular, at criticality the scaling can be consistent with a  non-unitary CFT with negative central charge~\cite{Renyi,nH_SSH}. This type of entropy definition is also related to pseudo entropy~\cite{PhysRevLett.130.031601,PhysRevD.103.026005}. At criticality, we find that the bipartite entropy scaling is consistent with $c=-2$ (See Appendix Sec.~\ref{app:nH_free_fermions} for a brief review of biorthogonal non-Hermitian physics and the non-Hermitian SSH model).

\def \tann{
	\begin{bmatrix}
		\psi_{+}(u)\\
		\psi_{-}(u)
\end{bmatrix}}
\def \pmann{
	\begin{bmatrix}
		\psi_{+}\\
		\psi_{-}
\end{bmatrix}}
\def \ann{
	\begin{bmatrix}
		\psi_{1}\\
		\psi_{2}
\end{bmatrix}}
\def \bann{
	\begin{bmatrix}
		\psi^{\dagger}_{1} & \psi^{\dagger}_{2}
\end{bmatrix}}
\def \tcre{
	\begin{bmatrix}
		\psi_{+}^\dagger(u)\\
		\psi_{-}^\dagger(u)
\end{bmatrix}}
\def \pmcre{
	\begin{bmatrix}
		\psi_{+}^\dagger\\
		\psi_{-}^\dagger
\end{bmatrix}}
\def \cre{
	\begin{bmatrix}
		\psi_{1}^\dagger\\
		\psi_{2}^\dagger
\end{bmatrix}}
\def\ketcre{
	\begin{bmatrix}
		\psi_+^\dagger(u) & \psi_-^\dagger(u)
\end{bmatrix}}

\subsection{Fermionic cMERA}
\label{subsec:fermionic_cMERA}

Here we review how to apply cMERA to a general two-band fermionic system and set up the operator evolution we will use for the non-Hermitian case. At half filling, the many-body ground state is a Gaussian (Slater) product
\begin{equation}
\ket{\Psi}=\prod_k \psi_-^\dagger(k)\ket{0},
\end{equation}
where $\psi_-^\dagger(k)$ creates the negative-energy single-particle eigenstate. Defining $\psi_+(k)$ as the annihilation operator for the positive-energy eigenstate, the ground state obeys
\begin{equation}
\psi_-^\dagger(k)\ket{\Psi}=0,\qquad \psi_+(k)\ket{\Psi}=0 \quad \text{for all } k.
\end{equation}

Instead of evolving the many-body state directly under cMERA, it is convenient to track the scale dependence of the creation and annihilation operators. A quadratic (bilinear) disentangler generates a linear transformation in the two-dimensional band space of each momentum sector. We define the right/left cMERA rotations (with $P$ and $\tilde{P}$ denoting path and reverse path ordering in $u$)
\begin{align}
    U_{R}(u) &= \tilde{P} \exp\!\Big(-i\int_{u}^{0}\tilde{K}(s)\,ds\Big), \\
    U_{L}(u) &= P \exp\!\Big(i\int_{u}^{0}\tilde{K}(s)\,ds\Big)
\end{align}
The operator flow from the UV ($u=0$) to a finite scale $u$ is
\begin{align}
	\tann
	&=U_R(u)\pmann_{\mathrm{UV}}U_L(u)
	= M(u)\,\ann,\\
	\tcre
	&=U_R(u)\pmcre_{\mathrm{UV}}U_L(u)
	= N(u)\,\cre.
\end{align}
The disentangler $\tilde{K}(u)$ is taken to be bilinear,
\begin{equation}
\tilde{K}(u)=\bann\,\mathcal{K}(u)\,\ann,
\end{equation}
with $\mathcal{K}(u)$ a $2\times2$ matrix acting in band space. Differentiating the annihilation sector with respect to $u$ gives
\begin{align}
\frac{d}{du}\tann
&=-iM(u)\comm{\tilde{K}(u)}{\ann} \nonumber\\
&=iM(u)\mathcal{K}(u)\ann,
\end{align}
so that
\begin{equation}
\frac{d}{du}M(u)=i\,M(u)\,\mathcal{K}(u).
\end{equation}
Repeating the procedure for the creation sector yields
\begin{equation}
\frac{d}{du}N(u)=-i\,N(u)\,\mathcal{K}^T(u).
\end{equation}
In the unitary case $\tilde{K}^\dagger=\tilde{K}$, one has $M^\dagger(u)=N^{\mathrm T}(u)$, so the flow preserves the usual Hermitian-adjoint relation between annihilation and creation operators, i.e.
\begin{equation}
    \big(\tann\big)^\dagger = \big(\tcre\big)^{\mathrm T}.
\end{equation}
In the non-unitary case this relation is absent; however, the biorthonormal contraction structure continues to hold beyond the unitary setup. To see this explicitly, consider the $2\times2$ contraction matrix in band space where $\expval{\dotsb}$ denotes the fermionic vacuum expectation value,
\begin{equation}
\begin{bmatrix}
\langle \psi_+^\dagger \psi_+ \rangle & \langle \psi_+^\dagger \psi_- \rangle\\
\langle \psi_-^\dagger \psi_+ \rangle & \langle \psi_-^\dagger \psi_- \rangle
\end{bmatrix}
= M(u)N^T(u).
\end{equation}
Using the flow equations, one finds that $M(u)N^T(u)$ is conserved along $u$, i.e.
\begin{align}
 	M(u)N^T(u)
 	&= M(0)U_L(u)U_R(u)N^T(0)\nonumber \\
 	&= M(0)N^T(0).
\end{align}
In the second equality, we used that $U_L(u)$ is the path-ordered inverse of $U_R(u)$. Thus even with a non-unitary cMERA circuit, the full $2\times2$ biorthonormal structure encoded in $\langle \psi_a^\dagger \psi_b\rangle$ is maintained during the flow.

Moreover, the rotated density matrix
\begin{equation}
\rho(u)=U_R(u)\,\rho\, U_L(u)
\end{equation}
generated by a non-unitary cMERA circuit thus automatically retains the left–right (biorthogonal) structure introduced above. This shows that cMERA is naturally consistent with biorthogonal formulations, and that the biorthogonal density matrix is the appropriate object for characterizing entanglement in a non-unitary CFT.

\subsection{Non-Unitary cMERA}

We now specialize the fermionic cMERA construction to the two-legged non-Hermitian SSH model. We adopt the standard cMERA ansatz in which the band-space rotation at momentum $k$ is parameterized by a single angle $\theta_k(u)$,
\begin{equation}
M(k,u)=
\begin{pmatrix}
\cos\theta_k(u) & \sin\theta_k(u)\\
-\sin\theta_k(u) & \cos\theta_k(u)
\end{pmatrix}.
\end{equation}
The boundary conditions are fixed by matching the UV state at $u=0$ and requiring the circuit to flow to a simple product reference state as $u\to-\infty$. Concretely,
\begin{equation}
\theta_k(u_{\rm IR}=-\infty)=\frac{\phi_{k=\pi}}{2},\qquad 
\theta_k(u_{\rm UV}=0)=\frac{\phi_k}{2},
\end{equation}
where $\phi_k$ is the complex band angle that characterizes the biorthogonal ground state of the non-Hermitian SSH model in Eq.~\eqref{eq:phase}.

For the disentangler we take a bilinear generator with the cutoff structure $\Gamma(|k|e^{-u}/\pi)$ introduced earlier, so that only modes within its support are disentangled at scale $u$. In each momentum sector we choose
\begin{equation}
i\mathcal{K}(k,u)\equiv G(k,u)=
\begin{pmatrix}
0 & -g(u)\\
g(u) & 0
\end{pmatrix}
\Gamma\!\left(\frac{|k|e^{-u}}{\pi}\right).
\end{equation}
It is worth noting that the kernel relevant for this lattice model is not of the usual continuum form $g\sim k$. The band phase to be disentangled in the SSH ground state is even under $k\to -k$, so the disentangler should also be even in $k$. On a lattice, a natural bilinear generator built from nearest-neighbor hopping carries factors $e^{\pm ik}$, whose even component is $\cos k$; in the long-wavelength limit this reduces to the momentum-independent form used in our ansatz ($\cos k\simeq 1$)~\footnote{
The momentum-independent term used in the long-wavelength cMERA ansatz corresponds to a local rotation in real space. If it were applied uniformly to a fixed lattice Hilbert space, it would not change the entanglement. In cMERA, however, the disentangler is applied together with a scale-dependent cutoff. As the RG coordinate \(u\) changes, modes outside the running cutoff are removed from the effective description. The local rotation can therefore perform the disentangling operation between the cutoff-removed sector and the remaining active modes, and hence can serve as a cMERA disentangler.
}.

The accumulated rotation from the UV down to a scale $u$ is determined by the integrated kernel,
\begin{equation}
\theta_k(0)-\theta_k(u)=\int_u^0 g(s)\,ds.
\label{eq:cmera_integral}
\end{equation}
Because the cutoff switches off the evolution of a fixed mode $k$ once it exits the support of $\Gamma$, the flow for that mode terminates at
\begin{equation}
u_{\rm IR}(k)=\ln(k/\pi).
\end{equation}
Hence the total rotation acquired by mode $k$ is fixed by the boundary values of $\theta_k$ and can be written as
\begin{equation}
\theta_k(0)-\theta_k(u_{\rm IR})
=\frac{\phi(k)}{2}-\frac{\phi(\pi)}{2}
=\int_{\ln(k/\pi)}^{0} g(s)\,ds .
\end{equation}
Since different momenta terminate at different depths $u_{\rm IR}(k)$, this relation determines the kernel layer by layer. Differentiating with respect to $k$ and using $k=\pi e^{u}$ gives
\begin{equation}
g(u)=-\frac{k}{2}\,\partial_k \phi(k)\Big|_{k=\pi e^{u}}.
\end{equation}
At criticality, the long-wavelength expansion of $\phi(k)$ around $k=0$ takes the logarithmic form
\begin{equation}
\phi(k)=-\frac{\pi}{2}+i\ln(ka)+\frac{i}{2}\ln\!\left(\frac{w}{4\mu}\right)+\mathcal{O}((ka)^2),
\end{equation}
which immediately implies a constant kernel in the infrared,
\begin{equation}
g(u)=-\frac{i}{2}.
\end{equation}
Thus, in the gapless long-wavelength regime the circuit approaches a scale-invariant non-unitary fixed point: $g(u)$ becomes a purely imaginary constant.

\subsection{The emergent spacetime}
\label{subsec:IIID_metric}

The emergent geometry is extracted from the information distance between neighboring cMERA layers. In a non-unitary circuit the bra is not fixed by Hermitian conjugation, so we define the RG metric component using the symmetric left–right overlap (a detailed discussion is provided in Appendix Sec.~\ref{app:LR_metric}):
\begin{eqnarray}
g_{uu}(u)du^2 = 1 - &&\braket{\Psi_L(u)}{\Psi_R(u+du)}\nonumber\\
&\times&\braket{\Psi_L(u+du)}{\Psi_R(u)}.
\end{eqnarray}
This reduces to the standard fidelity-based definition in the unitary limit $\bra{\Psi_L}=\bra{\Psi_R}$.

Since the cMERA state is Gaussian, the many-body overlap factorizes into independent momentum sectors, so $g_{uu}$ becomes an average of momentum-resolved contributions,
\begin{equation}
g_{uu}(u)
\;=\;
\frac{\int dk\, g_{ku}(u,k)}{\int dk},
\label{eq:50}
\end{equation}
with
\begin{eqnarray}
g_{ku}(u,k) du^2 = 1 - &&\braket{L_{k,-}(u)}{R_{k,-}(u+du)}\nonumber\\
&\times&\braket{L_{k,-}(u+du)}{R_{k,-}(u)}.
\end{eqnarray}
Expanding to leading order in $du$ yields
\begin{equation}
g_{ku}(u,k)
\;=\;
\big(\partial_u \theta_k(u)\big)^2.
\label{eq:52a}
\end{equation}
Using Eq.~\eqref{eq:cmera_integral} inside the active window, we obtain
\begin{equation}
g_{ku}(u,k)=g^2(u),
\qquad
g_{uu}(u)=g^2(u).
\label{eq:52}
\end{equation}
At the scale-invariant fixed point $g(u)=-\frac{i}{2}$, this yields
\begin{equation}
g_{uu}(u)=-\frac{1}{4},
\label{eq:53}
\end{equation}
in which the information metric has Lorentzian signature with $u$ timelike.

To complete the line element we incorporate the spatial rescaling generated by the dilation operator $L$, which implies $g_{xx}\propto e^{2u}$. Combining the two pieces gives
\begin{equation}
ds^2
\;=\;
-\frac{1}{4}\,du^2
\;+\;
\frac{\pi^2}{a^2}\,e^{2u}\,dx^2,
\label{eq:54}
\end{equation}
namely $(1+1)$-dimensional de Sitter space up to trivial rescalings.

Within the emergent de Sitter spacetime, it is useful to restate the cMERA construction in tensor-network language. Recall that the cMERA preparation of the UV density matrix can be written in a two-ended circuit form,
\begin{equation}
\vcenter{\hbox{\includegraphics[height=1.2cm]{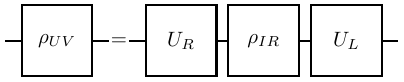}}}
\end{equation}
so that the UV density matrix is generated from a simple IR density matrix by a similarity transformation. Since the metric along the cMERA direction is timelike in the de Sitter metric extracted above, the two-ended circuit embeds naturally into the Penrose diagram (Fig.~\ref{fig:geo_pen}). The right and left halves of the cMERA circuit span the future and past planar patches of de Sitter space: the IR reference is coarse-grained to a single-site object at the center, while the two circuit endpoints extend to the asymptotic boundaries $\mathcal{I}^{\pm}$.

In summary, entanglement renormalization of the non-Hermitian two-legged SSH model yields an emergent de Sitter spacetime together with a Penrose-diagram embedding of the two-ended density-matrix circuit. In Appendix Sec.~\ref{app:other_free_fermions} we present a further example and show that both the non-Hermitian SSH model and the symplectic-fermion theory lead to the same emergent geometry.

\section{Ryu-Takayanagi surface in de Sitter space}
\label{sec:RT}

\subsection{De Sitter geodesic}

To establish a dS/cMERA correspondence in close analogy with the Hermitian AdS/cMERA case, we compare boundary entanglement scaling with the extremal length in the emergent geometry. For the metric obtained above, introduce the flat-slicing de Sitter line element
\begin{equation}
ds^2=-dt^2+\frac{\pi^2}{a^2}e^{2t/\alpha}dx^2,
\qquad
\alpha=\sqrt{|g_{uu}|},
\end{equation}
so that the geodesic-length functional can be written as
\begin{equation}
\int ds=\int dt\Bigl[-1+\frac{\pi^2}{a^2}e^{2t/\alpha}\dot x^{\,2}\Bigr]^{1/2}\equiv \int dt\,\mathcal L.
\label{eq:line_ele}
\end{equation}

Translational invariance in $x$ implies a conserved quantity. From the Euler--Lagrange equation one finds an integration constant $Q$ obeying
\begin{equation}
\frac{(\pi^2/a^2)e^{2t/\alpha}\dot x}{\Bigl[-1+(\pi^2/a^2)e^{2t/\alpha}\dot x^{\,2}\Bigr]^{1/2}}=Q.
\label{eq:Q}
\end{equation}
For a Lorentzian metric, the extremal curve relevant for the RT formula is naturally timelike, so the square root in the denominator of Eq.~\eqref{eq:line_ele} is purely imaginary. For $x(t)$ to be a physical spatial trajectory, $\dot{x}$ is real, and Eq.~\eqref{eq:Q} then requires the conserved constant $Q$ to be purely imaginary. We therefore write $Q=i|Q|$ and introduce the real parameter $\tilde Q\equiv a|Q|/\pi$, in terms of which
\begin{equation}
\dot x^{2}=\frac{a^2}{\pi^2}\frac{\tilde Q^{\,2}}{e^{4t/\alpha}+\tilde Q^{\,2}e^{2t/\alpha}}.
\label{eq:slope}
\end{equation}
Consistent with timelike causality, the extremal curve has no finite-$t$ turning point. Indeed, Eq.~\eqref{eq:slope} shows that $\dot x^{2}$ remains finite for all finite $t$, and can diverge only asymptotically as $t\to-\infty$. We therefore evaluate the geodesic that runs into the infinite past:
\begin{align}
\int ds&=\int_{t=0}^{t=-\infty}dt\Bigl[-1+\frac{\tilde Q^{\,2}e^{2t/\alpha}}{e^{4t/\alpha}+\tilde Q^{\,2}e^{2t/\alpha}}\Bigr]^{1/2}\nonumber \\
&=-i\alpha\sinh^{-1}(\tilde Q^{-1}).
\label{eq:geo_length}
\end{align}
The spatial projection is
\begin{align}
\frac{1}{a}\int dx&=\int_{t=0}^{t=-\infty}dt\,\frac{1}{\pi}\frac{\tilde Q}{\bigl(e^{4t/\alpha}+\tilde Q^{2}e^{2t/\alpha}\bigr)^{1/2}} \nonumber \\
&=\lim_{\epsilon\to0}\frac{\alpha}{\pi}\sqrt{\frac{1+\tilde Q^{2}}{\epsilon\,\tilde Q^{2}}}-\frac{\alpha}{\pi\tilde Q}\sqrt{1+\tilde Q^{2}},
\label{eq:spa_length}
\end{align}
which diverges as $\epsilon\to0$. At first glance this looks problematic if one tries to identify the boundary interval size with the spatial projection, $l=\int dx$. However, that identification is tied to the usual AdS/Hermitian setup where the extremal curve is anchored at the two endpoints of the boundary interval. In the present Lorentzian de Sitter geometry the relevant timelike geodesic does not return to the boundary, so the notion of interval size cannot be read off from a single bulk curve by $l=\int dx$. The divergence of $\int dx$ is therefore not a pathology; rather, it indicates that $l$ must be redefined.

\begin{figure}
	\includegraphics[width = 1.0\linewidth]{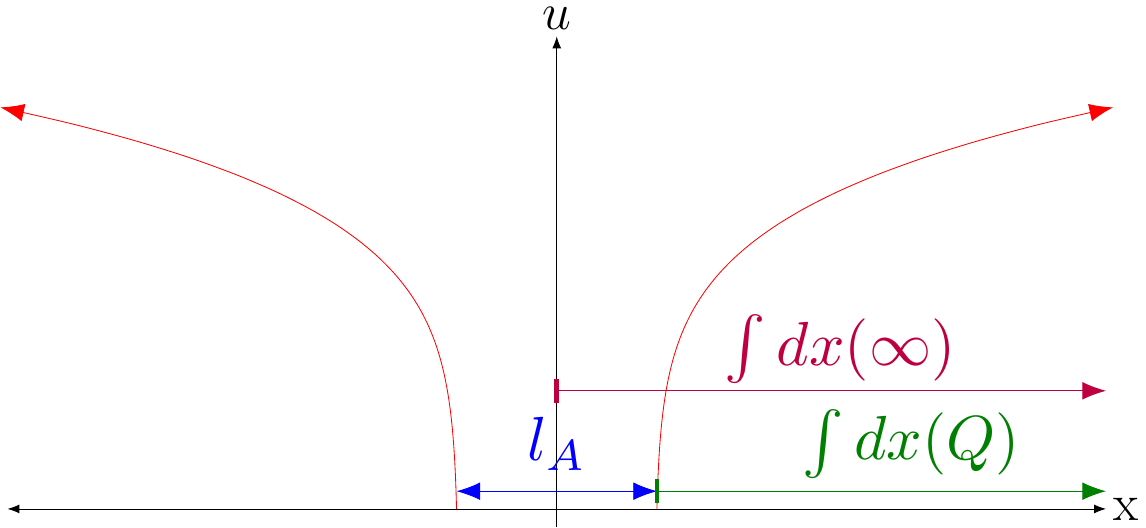}
	\caption{Outward-running de Sitter geodesic. In contrast to the AdS case, the extremal curve moves away from the boundary interval and converges to the common endpoint in the infinite past.}
    \label{fig:geo}
\end{figure}

We do so by matching the limiting behaviors of the extremal curve. When $\tilde Q\to0$, the solution approaches a vertical trajectory $x=\text{const}$ and the entropy tends to $S\to\int dt\,\sqrt{|g_{uu}|}$, appropriate to an infinite interval. When $\tilde Q\to\infty$, the curve becomes null-like and the proper length (hence the entropy) vanishes, $S\to0$, appropriate to a vanishing interval. These two limits motivate defining $l$ by subtracting the spatial projections relative to the $\tilde Q\to\infty$ reference curve,
\begin{equation}
l=\frac{2}{a}\int dx(\infty)-\frac{2}{a}\int dx(\tilde Q)
=\frac{2\alpha}{\pi\tilde Q}\sqrt{1+\tilde Q^{\,2}}-\frac{2\alpha}{\pi}.
\label{eq:interval}
\end{equation}
Geometrically, this definition implies that, unlike the AdS case, the extremal curve moves outward from the interval and all such curves converge to the same endpoint in the infinite past (Fig.~\ref{fig:geo}).

Eliminating $\tilde Q$ between Eq.~\eqref{eq:geo_length} and Eq.~\eqref{eq:interval} yields an explicit entropy--length relation,
\begin{equation}
S(l)=-i\,2\alpha\,\cosh^{-1}\Bigl(1+\frac{\pi l}{2\alpha}\Bigr),
\end{equation}
and for $l\gg1$,
\begin{equation}
S(l)\sim -i\,2\alpha\,\log\Bigl(\frac{\pi}{\alpha}\,l\Bigr).
\end{equation}
Thus, even in the flat-slicing Lorentzian de Sitter geometry---where extremal curves run into the asymptotic past---the extremal-length computation reproduces the expected logarithmic scaling at criticality. We will return to the interpretation of the overall proportionality coefficient in Sec.~\ref{sec:con}. For the massive theory, Appendix Sec.~\ref{app:massive} shows that the saturation entropy agrees with the corresponding vertical-geodesic cMERA prediction up to an additive constant, providing a gapped-case consistency check of the same entropy/geodesic matching.

\subsection{Connection to previous dS extremal surfaces}

RT extremal surfaces in de Sitter space have been discussed previously in dS holography setups where Lorentzian de Sitter is smoothly glued to a Euclidean cap~\cite{Hikida2022, PhysRevD.109.086009}.
In that capped geometry the endpoint of the surface is effectively fixed by the cap: on the Euclidean cap the real part of the geodesic/area functional is extremized at the poles, so boundary-anchored surfaces on a circle of total length $2\pi$ are driven to
\begin{equation}
\theta=\pm \frac{\pi}{2}\,,
\end{equation}
independent of the boundary interval.

Our flat-slicing construction can be viewed as the infinite-size limit of this situation.
If we rescale so that the global boundary circle has circumference $L$ and then take $L\to\infty$, the common endpoint $\theta=\pm\pi/2$ is pushed to the asymptotic end of the flat slice.
Equivalently, the statement ``all capped dS geodesics end at the poles'' becomes ``all flat-slicing geodesics end at the same point at infinity.''
In this sense, the outward-running geodesics and their convergence in the infinite past found above provide a flat-slicing viewpoint on the same de Sitter extremal-surface phenomenon.

\begin{figure}
	\includegraphics[width = 1.0\linewidth]{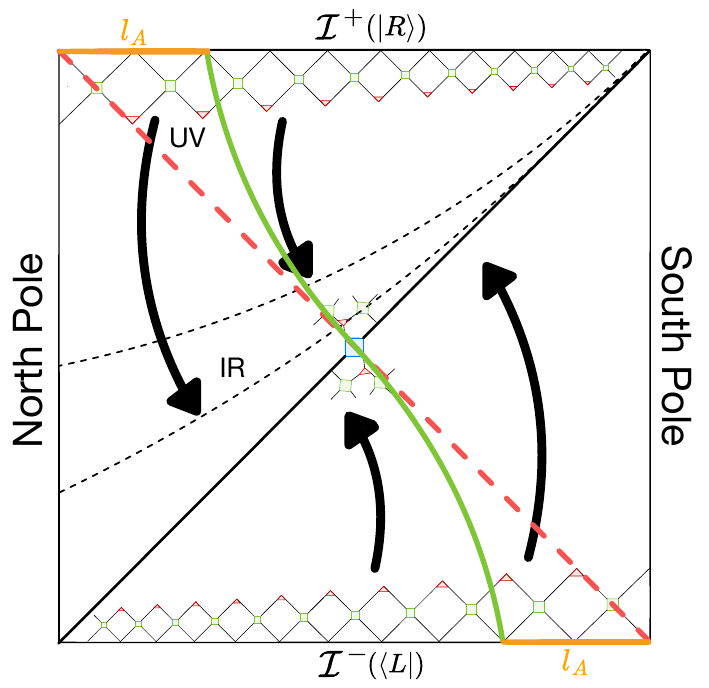}
	\caption{Tensor-network representation of the dS/cMERA construction and the associated geodesic cut. The right and left UV states are placed on the conformal boundaries \(\mathcal{I}^{\pm}\), while the RG flow terminates at an IR non-entangled non-Hermitian density matrix, shown in blue. Red triangles denote coarse-graining tensors and green boxes denote disentanglers. For a boundary subsystem \(A\), shown by the orange segment, the green curve indicates the corresponding de Sitter geodesic, which connects the interval endpoints on the right and left boundary states.}
    \label{fig:geo_pen}
\end{figure}

\subsection{Geometric lessons for the dS/MERA construction}
\label{subsec:tn_interpret}

The outward-running geodesics admit a direct interpretation in the tensor-network picture suggested by cMERA. Plotting the geodesic solutions on the Penrose-diagram embedding of the circuit (Fig.~\ref{fig:geo_pen}), each curve connects the endpoints of the subsystem on the right/left states placed at future and past infinity.
In the open-boundary setup considered here, the deep IR is coarse-grained to a single site at the center. Consequently, any curve (or cut) that connects the two asymptotic boundaries is funneled through this unique IR node, explaining why all geodesics pass through the same point.

Another useful intuition from Fig.~\ref{fig:geo_pen} comes from the zero-interval limit. As discussed above, when the subsystem size shrinks to zero the corresponding geodesic becomes null, giving the expected behavior $S\to 0$. From the tensor-network viewpoint, however, this may look less obvious at first glance: even when the boundary interval collapses to a point, the separating cut still extends toward the IR and appears to cross many legs.

In the tensor-network picture, this null limit is realized in a particularly simple way. For a vanishing boundary interval, the corresponding geodesic on the Penrose diagram follows the null ray shown in red in Fig.~\ref{fig:geo_pen}. Since the deep IR is coarse-grained to a single site at the center, it is natural to represent the coarse-graining of the open boundary by the direct connection from the boundary of the diagram to this central IR site, namely along the null ray. Along this boundary-null path, the cut lies on the boundary of the MERA network and intersects no tensor legs, so it carries zero bond-counting contribution. Hence, even though the cut extends all the way to the IR node, the vanishing interval yields zero entanglement entropy.

More generally, for a finite boundary interval, the green extremal surface in Fig.~\ref{fig:geo_pen} contains both a genuinely timelike segment and an approximately null continuation. Along the nearly null portion, the cut follows the network boundary and intersects no tensor legs, contributing zero bond-counting cost. The entanglement is therefore controlled by the part that deviates from the null direction. In this way, even though the extremal surface extends all the way toward the deep IR, the nonzero part of the geodesic follows the usual MERA scale organization: larger intervals are sensitive to longer-distance correlations encoded at deeper RG layers, whereas smaller intervals are controlled mainly by near-UV information.

\section{Connection to Discrete MERA and Bond-Counting Interpretation}
\label{sec:discrete_MERA}

In the previous sections, the emergent geometry was described in terms of continuum extremal surfaces. A central appeal of the MERA holographic picture, by contrast, lies in its discrete realization, where the tensor-network structure provides a microscopic interpretation of the geometry. In the AdS/MERA case, this discrete realization further leads to the bond-counting interpretation of the RT formula. The goal of this section is to construct the analogous bond-counting realization for dS/MERA.

In addition to the different endpoint structure---namely, that the relevant de Sitter extremal surface extends from the boundary endpoints toward the IR poles rather than directly between the two endpoints---the key requirement is inherited from the continuum result of Sec.~\ref{subsec:tn_interpret}. As the de Sitter extremal surface extends toward the IR pole, it undergoes a timelike-to-null transition. This transition preserves the usual MERA scale organization: although the surface reaches the deep IR, the nonzero length contribution remains controlled by the scale set by the boundary interval. A consistent dS/MERA representative should therefore mirror this transition in the discrete graph language.

We begin with the AdS/MERA construction, where assigning proper lengths to links in the tensor-suppressed MERA graph gives a discrete realization of the RT formula. This provides the benchmark for the dS/MERA bond-counting construction developed below.

\subsection{Review of bond-counting in AdS/MERA}

\begin{figure*}
    \includegraphics[width = 0.495\linewidth]{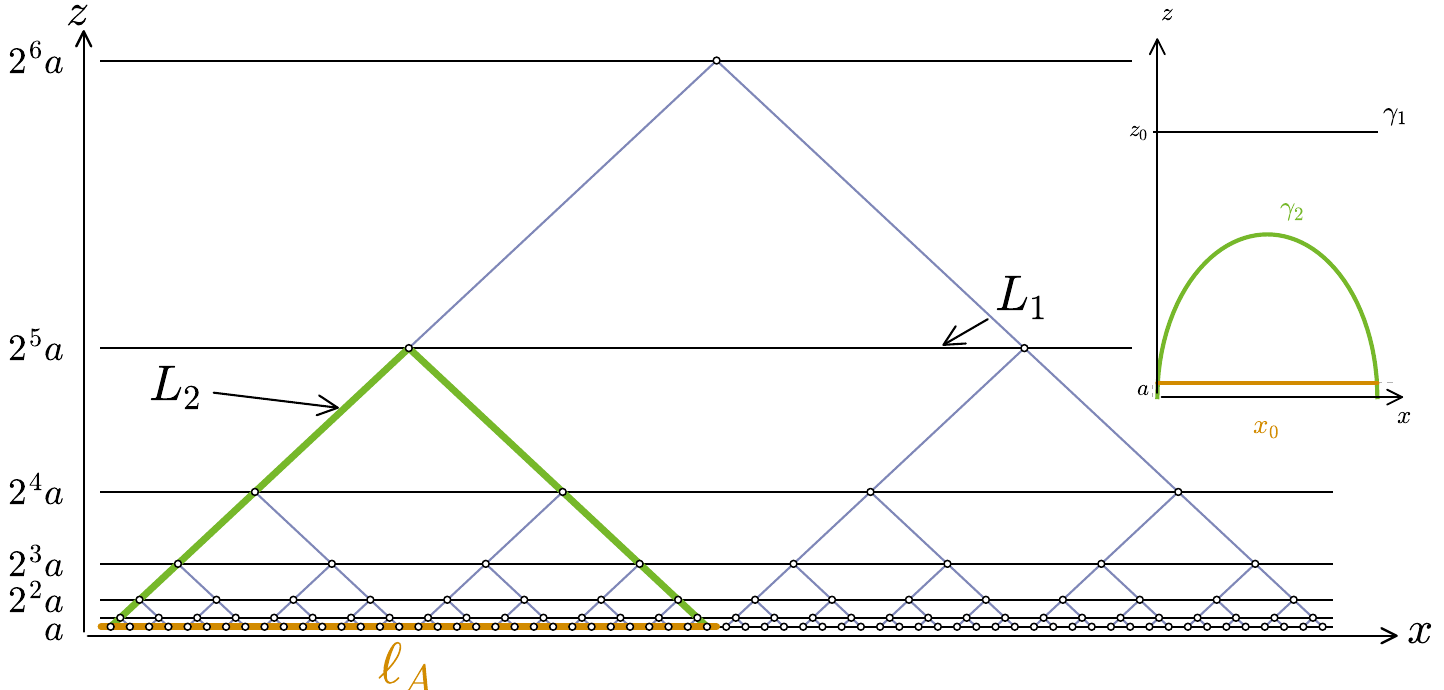}
    \includegraphics[width = 0.495\linewidth]{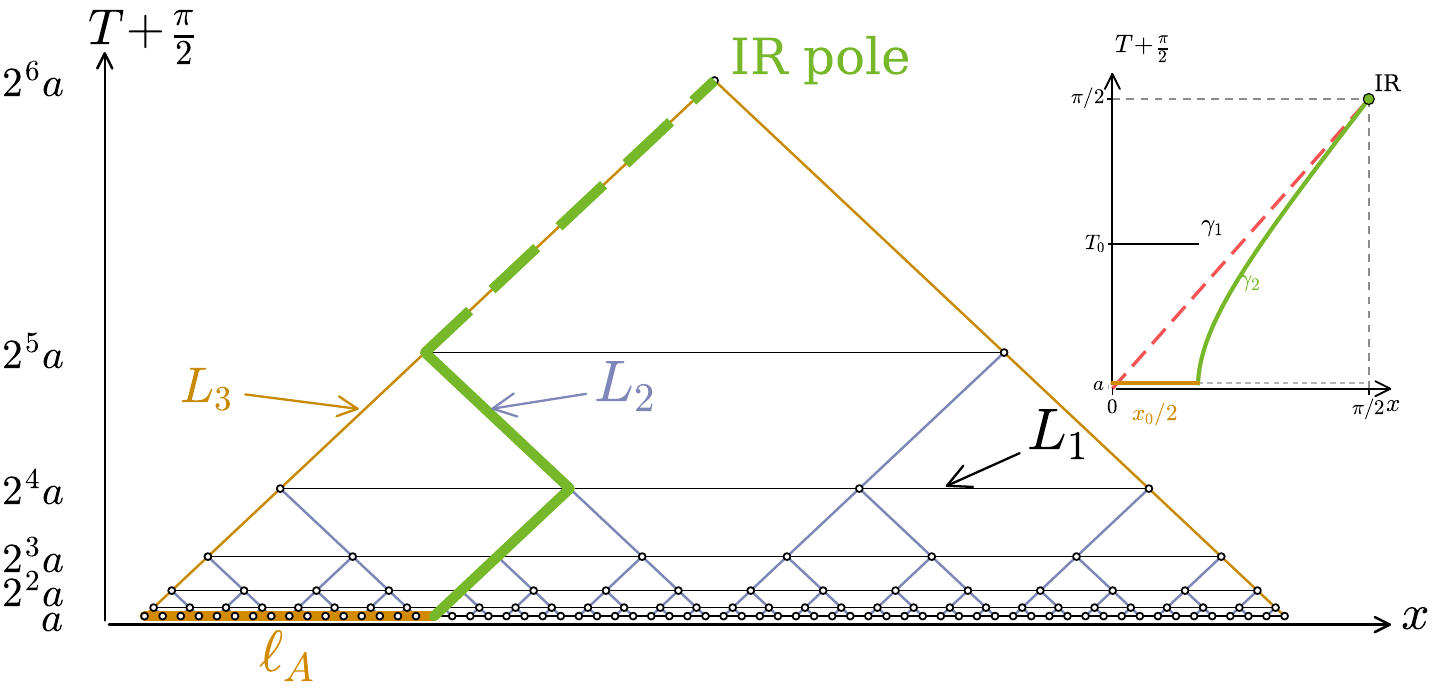} 
	\caption{Tensor-suppressed skeletons used for the bond-counting comparison. Circles denote effective sites at a given RG depth, horizontal links connect neighboring sites within the same layer, and diagonal links connect adjacent RG layers. The UV cutoff is \(a\), and the dyadic coarse-graining factor is chosen as \(k=2\).
[Left panel]: standard AdS/MERA skeleton with periodic boundary condition. The horizontal links carry cost \(L_1\), while the interlayer links carry cost \(L_2\). The boundary subsystem of size \(\ell_A\) is shown in orange, and the green path denotes the discrete representative of the AdS geodesic. The inset shows the corresponding continuum profiles on a constant-time slice of \(AdS\): \(\gamma_1\) is a horizontal profile at fixed \(z=z_0\), whereas \(\gamma_2\) is the boundary-anchored RT geodesic.
[Right panel]: dS/MERA skeleton with open boundary condition. In addition to the horizontal and interlayer costs \(L_1\) and \(L_2\), the boundary/ancilla links carry cost \(L_3\). The boundary subsystem of size \(\ell_A\) is again shown in orange, while the green path represents the discrete counterpart of the de Sitter extremal route toward the IR pole; the dashed green segment indicates its continuation along the null/boundary direction. The inset shows the corresponding continuum profiles in de Sitter Penrose coordinates: \(\gamma_1\) is a horizontal near-boundary profile, while \(\gamma_2\) is the extremal curve ending at the IR pole.}
    \label{fig:skeleton}
\end{figure*}

We begin by briefly reviewing the bond-counting construction of AdS/MERA~\cite{AdS_MERA_2}, since it provides the discrete template we will later generalize to the de Sitter case. The starting point is the MERA entropy bound in Eq.~\eqref{eq:mincut}, where the entropy of a boundary region \(A\) is controlled by the minimal cost of cutting bonds that separate \(A\) from its complement. To turn this cut problem into a geometric one, one suppresses the detailed tensors in the MERA and keeps only the graph formed by their connectivity. We refer to this reduced graph as the tensor-suppressed skeleton; see the left panel of Fig.~\ref{fig:skeleton}.

A cut \(\mathcal C_A\) in the original tensor network is then represented by a path \(\gamma\) on this skeleton. In this language, the weighted cut cost appearing in Eq.~\eqref{eq:mincut} is identified with the network length of the corresponding skeleton path,
\begin{equation}
\sum_i n_i(\mathcal C_A)\log\chi_i
\;\equiv\;
|\gamma|_{\rm net}.
\end{equation}
Here \(n_i(\mathcal C_A)\) counts the number of type-\(i\) bonds crossed by the cut, and \(\chi_i\) denotes the corresponding effective bond dimension.

For the scale-invariant MERA skeleton, circles denote effective sites at a given RG depth, horizontal links connect neighboring sites within the same layer, and vertical links connect adjacent RG layers. Translation and scale invariance imply that there are only two inequivalent bond types, so one assigns constant costs \(L_1\) and \(L_2\) to horizontal and vertical links, respectively. The network length is therefore written as
\begin{equation}
    |\gamma|_{\rm net}=L_1N_h(\gamma)+L_2N_v(\gamma),
\end{equation}
where \(N_h(\gamma)\) and \(N_v(\gamma)\) count the numbers of horizontal and vertical links along \(\gamma\), respectively.

To fix these costs, one matches the graph metric on the skeleton to the continuum geometry on a constant-time slice of \(AdS_3\),
\begin{equation}
ds^2=\frac{L^2}{z^2}(dz^2+dx^2),
\end{equation}
with UV cutoff \(z=a\). Consider the two representative curves shown in the inset of the left panel of Fig.~\ref{fig:skeleton}: a horizontal segment \(\gamma_1\) at fixed \(z=z_0\) spanning coordinate distance \(x_0\), and the boundary-anchored geodesic \(\gamma_2\) connecting two endpoints separated by the same \(x_0\). Their proper lengths are
\begin{equation}
|\gamma_1|_{\rm AdS}=\frac{L}{z_0}x_0,
\qquad
|\gamma_2|_{\rm AdS}=2L\ln\left(\frac{x_0}{a}\right).
\end{equation}

On the MERA skeleton shown in the left panel of Fig.~\ref{fig:skeleton}, the layer at depth \(m\) is identified with \(z_0=k^m a\). Here $k$ is the RG reduction of the effective interval size. A horizontal skeleton path on that layer contains \(x_0/(k^m a)=x_0/z_0\) horizontal links, so
\begin{equation}
|\gamma_1|_{\rm MERA}=L_1\frac{x_0}{z_0}.
\end{equation}
Matching \( |\gamma_1|_{\rm MERA}=|\gamma_1|_{\rm AdS} \) fixes the horizontal link cost to be \(L_1=L\). For two UV sites separated by \(x_0\gg a\), the shortest skeleton path is the usual up--down trajectory shown by the green curve in the left panel of Fig.~\ref{fig:skeleton}: after \(\log_k(x_0/a)\) RG steps, the separation is reduced to order one, so the path contains \(2\log_k(x_0/a)\) vertical links and hence
\begin{equation}
|\gamma_2|_{\rm MERA}=2L_2\log_k\left(\frac{x_0}{a}\right).
\end{equation}
Matching \( |\gamma_2|_{\rm MERA}=|\gamma_2|_{\rm AdS} \) then gives $L_2=L\ln k$. With these weights fixed, the discrete RT rule becomes transparent. A horizontal path has length of order \(x_0/a\), whereas the up--down MERA geodesic grows only logarithmically. The latter therefore reproduces the CFT scaling 
\begin{equation}
    S_A \sim |\gamma_2|_{\rm MERA} = 2L\ln\left(\frac{x_0}{a}\right),
\end{equation}
giving the bond-counting realization of the RT formula in AdS/MERA.

\subsection{Bond counting in dS/MERA}

We now extend this AdS/MERA skeleton bond-counting logic to dS/MERA using the geometric lessons of Sec.~\ref{subsec:tn_interpret}. The first lesson concerns the global structure of the extremal surface. In the open-boundary-condition (OBC) construction, the extremal surface connects the paired right/left endpoints at future and past infinity and is funneled through the coarse-grained IR node. Thus the discrete cut should not be modeled as an AdS-like surface that turns back within a single side, but as a cut that continues toward the IR node of the two-ended skeleton. The second lesson comes from the zero-interval limit. In this limit, the extremal surface becomes a null ray along the boundary of the MERA network. Along this boundary segment, the cut lies outside the tensor network and intersects zero tensor legs. The corresponding link in the skeleton graph is therefore assigned zero bond-counting cost. We encode this boundary continuation by a third skeleton cost \(L_3=0\). The network cost of a skeleton path \(\gamma\) is then defined by the weighted link count
\begin{equation}
    |\gamma|_{\rm net}=L_1N_h(\gamma)+L_2N_v(\gamma)+L_3N_n(\gamma),
\end{equation}
where \(N_h,N_v\) and \(N_n\) count the numbers of horizontal, vertical, and null links along \(\gamma\), respectively.

With this skeleton rule in place, the link costs can be fixed by matching to the continuum de Sitter lengths, in direct parallel with the AdS/MERA construction. The emergent de Sitter spacetime can be written in conformal coordinates as
\begin{equation}
ds^2
=
\frac{L^2}{\cos^2 T}
\left(
-dT^2+dx^2
\right),
\end{equation}
with UV cutoff \(T=-\pi/2+a\). Consider the two representative profiles shown in the inset of the right panel of Fig.~\ref{fig:skeleton}: a horizontal segment \(\gamma_1\) at fixed \(T=-\pi/2+T_0\) spanning a coordinate distance \(x_0/2\) from the open boundary, and the extremal surface \(\gamma_2\) connecting the endpoint to the IR pole with the same boundary separation \(x_0/2\).
\begin{align}
|\gamma_1|_{\rm dS}
&=
\frac{L}{\cos(-\pi/2+T_0)}
\frac{x_0}{2}
\simeq
\frac{L}{2T_0}x_0,
\\
|\gamma_2|_{\rm dS}
&=
iL\sinh^{-1}
\left(
\sin\frac{x_0}{2}\cot a
\right)
\simeq
iL\ln\left(\frac{x_0}{a}\right).
\end{align}
In the last expressions we used \(T_0\ll 1\), \(a\ll 1\), \(x_0\ll 1\), and \(x_0/a\gg 1\).

On the dS/MERA skeleton shown in the right panel of Fig.~\ref{fig:skeleton}, the layer at depth \(m\) is identified with \(T_0=k^m a\). For the binary construction considered here, \(k=2\). A horizontal skeleton path on that layer contains \(x_0/(2k^m a)=x_0/(2T_0)\) horizontal links, so
\begin{equation}
|\gamma_1|_{\rm MERA}
=
L_1\frac{x_0}{2T_0}.
\end{equation}
Matching \( |\gamma_1|_{\rm MERA}=|\gamma_1|_{\rm dS} \) fixes the proper length assigned to each horizontal link, \(L_1=L\).

We now describe the skeleton representative of the continuum extremal surface \(\gamma_2\), shown as the green path in the right panel of Fig.~\ref{fig:skeleton}. For definiteness, take an OBC boundary interval \(A=[0,x_0/2]\) with \(x_0\gg a\). Starting from the interior endpoint of \(A\), the green path first moves through the tensor-network bulk, while coarse graining reduces its effective distance to the open boundary. When this distance becomes of order one lattice spacing, the same path continues to the IR pole along the boundary-null segment. Thus only the bulk part of the green path contributes to the nonzero skeleton length. Since each RG layer reduces the effective interval size by a factor \(k\), this bulk part contains \(\log_k(x_0/a)\) RG links, hence
\begin{equation}
|\gamma_2|_{\rm MERA}
=
L_2\log_k\left(\frac{x_0}{a}\right).
\end{equation}
This gives the desired logarithmic scaling of the skeleton proper length. Matching \( |\gamma_2|_{\rm MERA}=|\gamma_2|_{\rm dS} \) fixes the proper length assigned to each RG link, $L_2=iL\ln k$. The continuum timelike-to-null transition is thus mirrored in the skeleton graph by a discrete transition from vertical links $L_2$ to null links $L_3$.

\begin{figure*}
    \includegraphics[width = .5\linewidth]{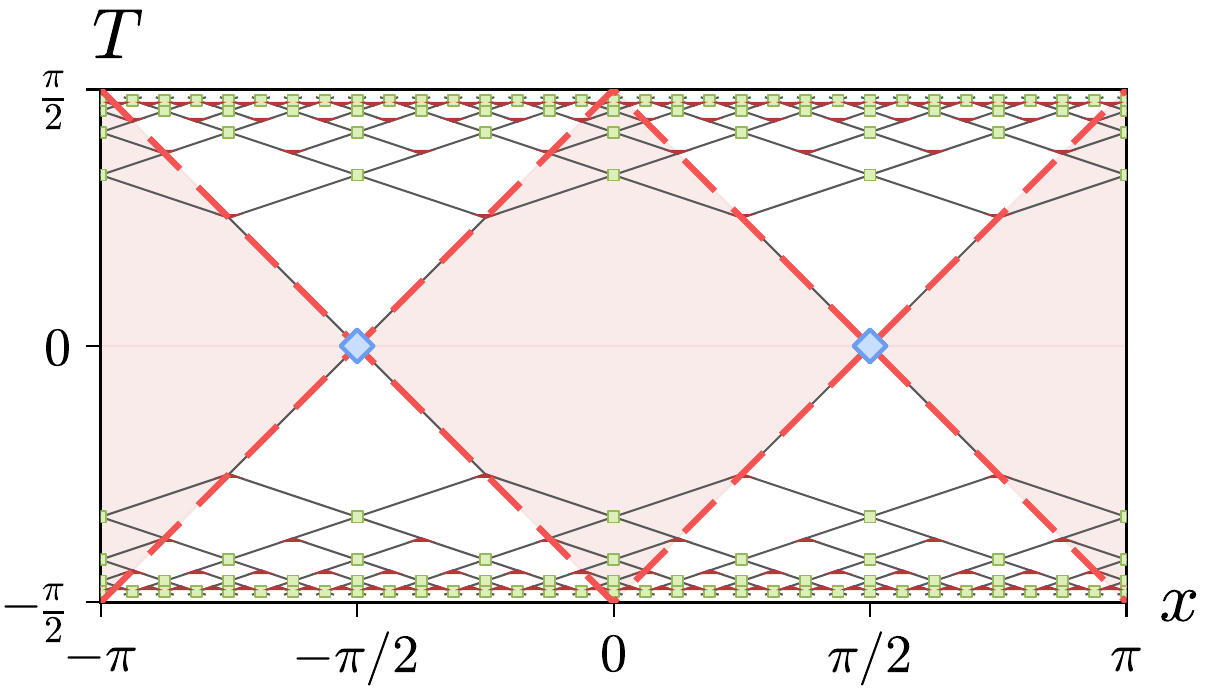} 
    \includegraphics[width = 0.49\linewidth]{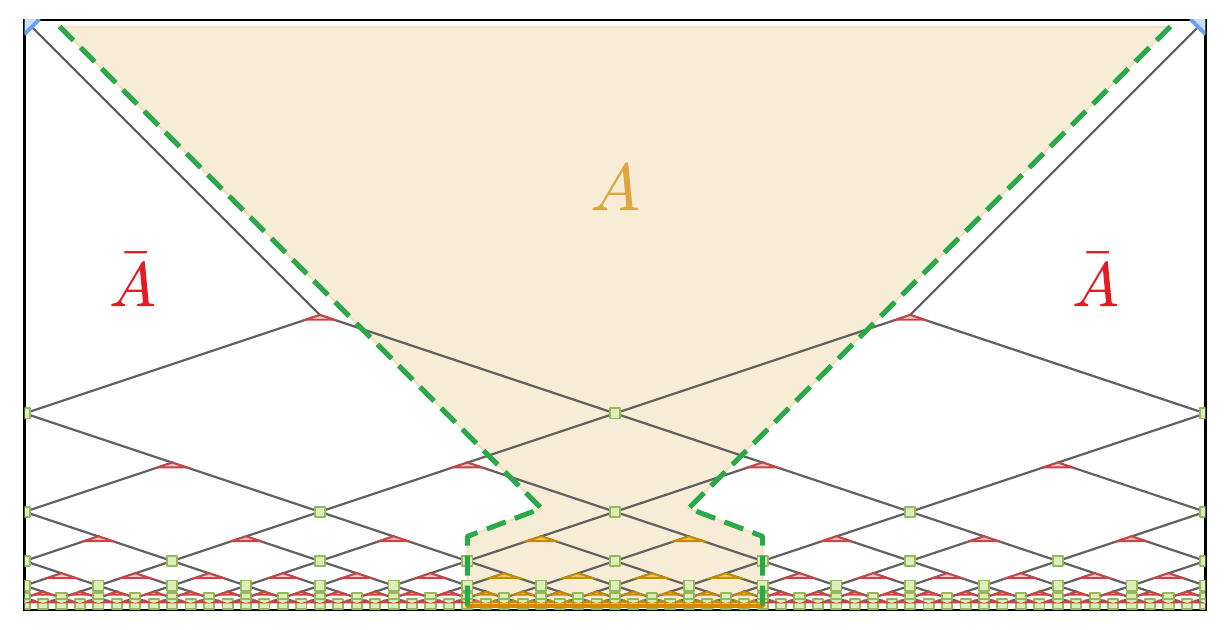} 
    \caption{
    Periodic-boundary-condition tensor-network construction in the de Sitter Penrose diagram and local \(A|\bar A\) assignment. Triangles denote coarse-graining tensors, square boxes denote entanglers, and blue diamonds denote the IR reference density-matrix sites.
[Left panel]: global two-ended tensor-network representation in Penrose coordinates. The dashed red lines indicate the null horizons.
[Right panel]: local enlargement of the region \(x\in[-\pi/2,\pi/2]\), \(T\in[-\pi/2,0]\), using the same fraction of the boundary interval as the right panel of Fig.~\ref{fig:skeleton}. In this panel, the \(A|\bar A\) assignment is indicated by the color of the coarse-graining tensors: yellow for \(A\) and red for \(\bar A\). The shaded yellow region denotes \(A\), while the exterior regions denote \(\bar A\). The dashed green cut first follows the timelike direction and then continues along the null direction. This construction contains one effective null link, represented by the top entangler cut. Although the cut visually passes through all legs of the top entangler, all these indices are assigned to the same \(\bar A\) group. Hence the null continuation does not sever an \(A|\bar A\) bond and contributes zero effective cost.
    }
    \label{fig:pbc_tensor}
\end{figure*}

The resulting link cost is complex-valued, $L_2 =i L \ln k$. In standard Hermitian architectures, graph edges are strictly real and positive to denote physical bond capacities $\log \chi$, mapping directly to positive-definite entanglement bounds. In non-unitary tensor networks, the cost of a link can be complex, and the entropic bound in Eq.~\eqref{eq:mincut} no longer holds. Nevertheless, the extremal cut connecting the boundary interval endpoint to the deep-IR pole still reproduces the logarithmic scaling of the non-unitary entanglement entropy,
\begin{equation}
S_A\sim|\gamma_2|_{\rm MERA}=iL\ln\left(\frac{x_0}{a}\right),
\end{equation}
which provides a possible bond-counting picture of the de Sitter RT formula.

Having established the OBC skeleton bond-counting rule, we now turn to the periodic-boundary-condition (PBC) setup. The PBC case requires a more careful implementation of the zero-cost \(L_3=0\) link. In the OBC skeleton, the boundary-null part of the cut can run outside the tensor network and therefore intersects zero tensor legs. In a system with PBC, however, there is no open boundary along which this exterior continuation can literally run. Nevertheless, in the PBC tensor-network construction introduced below, an effective zero-cost null link can still be realized. Below we show, by examining the tensor-network realization of the corresponding \(A|\bar A\) cut, that this null continuation severs no tensor legs and therefore has zero bond-counting cost.

To construct the PBC network, we adopt the Penrose-diagram coordinates and dyadic layer structure used in the de Sitter tensor-network construction of Ref.~\cite{PhysRevD.105.125009}. More explicitly, we use conformal coordinates \( (x,T) \) on \( dS_2 \), with \( x\sim x+2\pi \) and \( -\pi/2<T<\pi/2 \). We place the \( n \)-th layer at
\begin{equation}
T_n^{(\pm)}
=
\pm \alpha_n,
\qquad
\alpha_n
=
\frac{\pi}{2}\left(1-2^{-n}\right),
\end{equation}
where \( + \) and \( - \) denote the future and past halves of the two-ended tensor network.

For an interval centered at \( x_c \), we introduce the shifted angular coordinate $\tilde x=x-x_c$ with $\tilde x\in[-\pi,\pi)$. The tensor-network sites on the \( n \)-th layer are placed at the dyadic midpoint positions
\begin{equation}
\tilde x_{n,j}
=
-\pi+\frac{\pi}{2^n}\left(j+\frac{1}{2}\right),
\qquad
j=0,1,\cdots,2^{n+1}-1,
\end{equation}
as illustrated in the left panel of Fig.~\ref{fig:pbc_tensor}. With this site placement, the number of sites inside the pink static patch remains unchanged along the RG direction, in the same spirit as the stationary causal cone discussed in Ref.~\cite{PhysRevD.96.123536}. In fact, no tensor-network site lies inside the chosen static patch. This property is the key ingredient that allows the null continuation of the extremal-surface cut to acquire zero bond-counting cost in the PBC skeleton.

We now explain how the zero-cost null link is realized in the PBC tensor network, as illustrated in the right panel of Fig.~\ref{fig:pbc_tensor}. The cut associated with the extremal surface of a single interval \(A\) defines an \(A|\bar A\) assignment of the tensor-network sites. In the local enlargement, the cut starts from the two endpoints of \(A\), first follows the timelike direction through the tensor-network bulk, and then connects to the null continuation. Along the timelike part, the cut separates sites assigned to \(A\) from sites assigned to \(\bar A\), and therefore severs tensor legs; this part is represented in the skeleton by links with nonzero bond-counting cost. The null part is different. Although the dashed green cut visually passes through all legs of the top entangler, this visual crossing does not by itself determine the bond count. The relevant question is whether the tensor indices of the entangler are split by the \(A|\bar A\) separation. In the assignment shown in the figure, all indices of the top entangler belong to the same \(\bar A\) group. The null continuation therefore severs no tensor legs between \(A\) and \(\bar A\), and the corresponding skeleton link is assigned \(L_3=0\). This is how the PBC construction realizes the zero-cost null link even though there is no exterior open boundary.

Therefore, the skeleton representative of the PBC extremal surface is effectively a doubled version of the OBC construction. The two UV branches give the logarithmic bond-counting contribution, while the two null continuations to the IR poles are assigned \(L_3=0\). In this way, the PBC tensor network realizes the bond-counting counterpart of the de Sitter extremal surface and reproduces the desired logarithmic scaling.

\section{Discussion and conclusion}
\label{sec:con}

Tensor-network constructions have long provided a useful language for holographic entanglement. In the AdS setting, MERA gives a discrete realization of RT-type cut geometry, while cMERA provides a continuum entanglement-renormalization route to an emergent bulk metric. In this work, we developed a de Sitter analogue of this AdS/(c)MERA structure in a concrete non-Hermitian critical system. The continuum part of the construction connects the microscopic non-Hermitian boundary state to an emergent de Sitter geometry, while the discrete part uses the resulting Penrose-diagram structure to organize a MERA tensor network and its extremal-cut interpretation.

On the continuum side, we constructed a non-unitary cMERA circuit for a non-Hermitian critical fermion chain, viewed as a lattice realization of a non-unitary CFT. The non-Hermitian entangler naturally leads to a biorthogonal left--right organization of the circuit, providing an appropriate in/out pairing for overlaps and correlators in the non-unitary setting. From the local entangling data of this circuit, we extracted an effective bulk metric whose RG direction is timelike. At criticality, this metric becomes \((1+1)\)-dimensional de Sitter space. This gives a microscopic entanglement-RG realization of de Sitter kinematics in a non-unitary critical theory.

Motivated by this continuum emergence of de Sitter geometry, we then developed a discrete MERA construction adapted to the de Sitter Penrose diagram. The relevant de Sitter extremal curve differs qualitatively from the AdS RT geodesic: it bends toward spatial infinity, different extremals approach the same asymptotic region, and the continuation is past-directed. These features are naturally reflected in the Penrose-embedded tensor network. The coarse-graining structure drives boundary data toward the deep-IR pole, while the two-ended biorthogonal organization gives a natural tensor-network interpretation of the future--past connection across the diagram.

We further formulated a de Sitter version of the AdS/MERA bond-counting picture. For this purpose, we pass from the full tensor network to its tensor-suppressed skeleton, which keeps only the connectivity relevant for path-cost computations. The relevant extremal cut connects the boundary interval endpoint to the deep-IR pole, and its null continuation is represented by zero-cost links because it crosses no tensor legs. With these rules, the skeleton path reproduces the timelike-to-null structure of the de Sitter extremal surface and gives the expected logarithmic dependence of the entropy. Since the corresponding link cost is complex, it is not naturally interpreted as a positive bond capacity \(\log\chi\). The usual entropy-bound interpretation of Hermitian MERA is therefore not available in this form. Nevertheless, the construction provides a possible bond-counting picture for the de Sitter extremal surface.

The PBC tensor network may also suggest a special encoding structure if entanglement-wedge reconstruction still holds for de Sitter holography. In the Penrose-embedded tensor network, the corresponding static patch contains no explicit tensor-network sites, suggesting that a boundary interval is not naturally associated with independent degrees of freedom inside the corresponding static patch. This structure is consistent with the continuum extremal-surface picture: in the vanishing-interval limit, the extremal profile becomes a null geodesic along the static-patch horizon, and a vanishing boundary interval should not contain bulk information. This points to a possible special feature of de Sitter tensor networks, whose reconstruction meaning remains an interesting question for future work.

We turn to a notable feature of our extremal-surface comparison: the coefficient phase mismatch. The issue here is not merely an overall proportionality ambiguity, which is common in cMERA frameworks, but a structural phase difference. Specifically, the timelike de Sitter geodesic yields a purely imaginary geometric coefficient, $c_{\rm geo}= - 3 i$, whereas the microscopic non-Hermitian critical chain features a real, negative effective central charge, $c=-2$. In our construction, the agreement in logarithmic scaling, the emergent \(dS_2\) geometry from cMERA, and the timelike-to-null structure in dS/MERA suggest that the relation is unlikely to be accidental. This mismatch may instead indicate an additional dictionary factor between boundary biorthogonal entropy and bulk extremal-surface length in a possible dS/CFT RT formula.

More broadly, our construction suggests a bottom-up route toward de Sitter tensor networks. Instead of starting from a purely geometric tensor-network ansatz, one can first extract the continuum spacetime structure from boundary information-RG data and then use its causal and extremal-surface properties to constrain the discrete network. It would be useful to explore this circuit-guided logic in  non-unitary tensor-network models where bond-counting rules are more directly realized, such as HaPPY-code or random-tensor-network constructions, and to extend the present \((1+1)\)-dimensional framework to higher-dimensional de Sitter extremal surfaces. 
Testing the boundary limits of this extended framework provides an exciting avenue for future work. We hope that this combined continuum and discrete construction provides a useful step toward a more concrete tensor-network understanding of de Sitter holography.

\section{Acknowledgements}
This work was supported by National Science and Technology Council of Taiwan under Grants No. NSTC 113-2112-M-007-019, 114-2918-I-007-015. P.-Y.C thanks Shinsei Ryu and Dong-han Yeom for stimulating discussions and acknowledges support from the National Center for Theoretical Sciences, Physics Division.
We also thank RIKEN iTHEMS, where this work benefited from stimulating discussion during the workshop on "de Sitter Holography Meets Non-Hermitian Quantum Matter".  

\appendix 

\section{Biorthogonal density matrices and representative non-Hermitian critical free-fermion systems}
\label{app:nH_free_fermions}

In this appendix we review the biorthogonal density-matrix formalism used in the main text and summarize two representative non-Hermitian critical free-fermion systems: the non-Hermitian SSH model and the lattice symplectic-fermion model. 

\subsection{Biorthogonal Gaussian states and correlation-matrix entropy}
\label{subsec:biorthogonal_correlation_matrix}

We consider a generic quadratic non-Hermitian fermion Hamiltonian
\begin{equation}
H=\sum_{ij}c_i^\dagger \mathcal{H}_{ij}c_j,
\qquad
\mathcal{H}^\dagger\neq \mathcal{H}.
\end{equation}
The right and left single-particle eigenstates are defined by
\begin{equation}
\mathcal{H}|R_\alpha\rangle=\epsilon_\alpha |R_\alpha\rangle,
\qquad
\mathcal{H}^\dagger |L_\alpha\rangle=\epsilon_\alpha^* |L_\alpha\rangle.
\end{equation}
With the biorthogonal normalization $\langle L_\alpha|R_\beta\rangle=\delta_{\alpha\beta}$, the single-particle Hamiltonian has the spectral representation
\begin{equation}
\mathcal{H}
=
\sum_\alpha
\epsilon_\alpha
|R_\alpha\rangle\langle L_\alpha|.
\end{equation}
The corresponding many-body Hamiltonian can be written as
\begin{equation}
H
=
\sum_\alpha
\epsilon_\alpha
\psi^\dagger_{R,\alpha}\psi_{L,\alpha},
\qquad
\{\psi_{L,\alpha},\psi^\dagger_{R,\beta}\}
=
\delta_{\alpha\beta}.
\end{equation}

For a filled Fermi sea, the right and left many-body ground states are
\begin{equation}
|G_R\rangle
=
\prod_{\alpha\in \mathrm{occ}}
\psi^\dagger_{R,\alpha}|0\rangle,
\qquad
\langle G_L|
=
\langle 0|
\prod_{\alpha\in \mathrm{occ}}
\psi_{L,\alpha},
\end{equation}
with $\langle G_L|G_R\rangle=1$. The density matrix associated with this biorthogonal pure state is
\begin{equation}
\rho=|G_R\rangle\langle G_L|.
\end{equation}
It is generally non-Hermitian, \(\rho^\dagger\neq \rho\), but remains normalized and projective, \(\mathrm{Tr}\rho=1\) and \(\rho^2=\rho\). For a spatial bipartition \(A\cup \bar A\), the reduced density matrix is defined by
\begin{equation}
\rho_A=\mathrm{Tr}_{\bar A}\rho,
\qquad
S_A=-\mathrm{Tr}_A\,\rho_A\ln\rho_A.
\end{equation}
Thus the entropy is computed from \(\rho_A=\mathrm{Tr}_{\bar A}|G_R\rangle\langle G_L|\), rather than from the Hermitian density matrix \(|G_R\rangle\langle G_R|\).

For a quadratic free-fermion system, the biorthogonal ground state is Gaussian. The reduced density matrix on \(A\) can therefore be written as
\begin{equation}
\rho_A
=
Z^{-1}e^{-\mathcal{H}_E},
\qquad
\mathcal{H}_E
=
\sum_{i,j\in A}
h^E_{ij}c_i^\dagger c_j .
\end{equation}
The correlation matrix on \(A\) is defined in trace form by
\begin{equation}
C_{ij}
=
\langle G_L|c_i^\dagger c_j|G_R\rangle
=
\mathrm{Tr}\left(\rho_A c_i^\dagger c_j\right),
\qquad
i,j\in A.
\end{equation}
Equivalently, it is the restriction of the occupied-band projector to the subsystem \(A\),
\begin{equation}
C_{ij}
=
\sum_{\alpha\in \mathrm{occ}}
L_{\alpha i}R^\dagger_{\alpha j}.
\end{equation}

To relate \(C_A\) to the entropy, we diagonalize the single-particle entanglement Hamiltonian by a similarity transformation,
\begin{equation}
h^E
=
S\,\mathrm{diag}(\epsilon_n)\,S^{-1}.
\end{equation}
Writing \(c=(c_1,\ldots,c_{|A|})^T\), the corresponding biorthogonal entanglement modes may be chosen as
\begin{equation}
f_L=S^{-1}c,
\quad
f_R^\dagger=c^\dagger S,
\quad
\{f_{L,m},f^\dagger_{R,n}\}=\delta_{mn}.
\end{equation}
In this basis,
\begin{equation}
\mathcal{H}_E
=
\sum_n
\epsilon_n f^\dagger_{R,n}f_{L,n},
\quad
\rho_A
=
\prod_n
\frac{
e^{-\epsilon_n f^\dagger_{R,n}f_{L,n}}
}{
1+e^{-\epsilon_n}
}.
\end{equation}
The single-mode occupations are therefore
\begin{equation}
\nu_n
=
\langle f^\dagger_{R,n}f_{L,n}\rangle
=
\frac{1}{1+e^{\epsilon_n}}.
\end{equation}
The von Neumann entropy can then be written as
\begin{align}
S_A
&=
-\mathrm{Tr}\left(\rho_A\ln\rho_A\right)
\nonumber\\
&=
-\mathrm{Tr}
\left[
\frac{e^{-h^E}}{1+e^{-h^E}}
\ln
\frac{e^{-h^E}}{1+e^{-h^E}}
+
\frac{1}{1+e^{-h^E}}
\ln
\frac{1}{1+e^{-h^E}}
\right]
\nonumber\\
&=
-\sum_n
\left[
\nu_n\ln\nu_n
+
(1-\nu_n)\ln(1-\nu_n)
\right].
\label{eq:app_entropy_correlation_formula}
\end{align}
Equivalently,
\begin{equation}
S_A
=
-\mathrm{Tr}
\left[
C_A\ln C_A
+
(1-C_A)\ln(1-C_A)
\right].
\end{equation}
This is the expression used in the numerical calculations below and in the main text. Since \(C_A\) is generally non-Hermitian, its eigenvalues are not constrained to the interval \([0,1]\), and the corresponding biorthogonal entropy may be negative or complex.

\subsection{Non-Hermitian SSH model and symplectic fermion}
\label{subsec:nhssh_and_symplectic}

We now introduce the two representative lattice models considered in this appendix. For each model, the biorthogonal entanglement entropy is computed from Eq.~\eqref{eq:app_entropy_correlation_formula} and fitted to the critical logarithmic form.

\paragraph{Non-Hermitian SSH model.}

We first introduce the non-Hermitian SSH model discussed in Ref.~\cite{nH_SSH}, which is a two-band free-fermion model with balanced gain and loss. We write its Bloch Hamiltonian as
\begin{equation}
\mathcal{H}_{\mathrm{SSH}}(k)
=
\begin{pmatrix}
iu & q(k) \\
q^*(k) & -iu
\end{pmatrix},
\quad
q(k)=v-w e^{-ik},
\label{eq:app_nhssh_hamiltonian}
\end{equation}
where \(u,v,w\in\mathbb{R}\). The spectrum is
\begin{equation}
E_\pm(k)
=
\pm
\sqrt{|q(k)|^2-u^2}.
\label{eq:app_nhssh_spectrum}
\end{equation}
The exceptional critical point occurs when the minimum of \(|q(k)|\) reaches \(|u|\). For the real-parameter convention used here, this gives
\begin{equation}
|v|-|w|=\pm|u|,
\label{eq:app_nhssh_critical_condition}
\end{equation}
which contains both the trivial critical line with $|v|>|w|$ and the topological critical line with $|w|>|v|$. The entanglement scaling on the topological critical line was further discussed in Ref.~\cite{chou2026ptsymmetryenrichednonunitarycriticality}.

The two-legged model used in the main text can be written as a symmetric hopping convention of the same SSH-type structure. Replacing the hopping factor by its mirror-symmetric combination gives
\begin{equation}
q(k)
=
v-\frac{w}{2}\left(e^{ik}+e^{-ik}\right)
=
v-w\cos k .
\label{eq:app_two_leg_qk}
\end{equation}
It has the same spectral form as Eq.~\eqref{eq:app_nhssh_spectrum} and satisfies the positive-sign branch of the critical condition in Eq.~\eqref{eq:app_nhssh_critical_condition}.

\begin{figure}[t]
  \centering
  \begin{minipage}[b]{0.48\linewidth}
    \panel{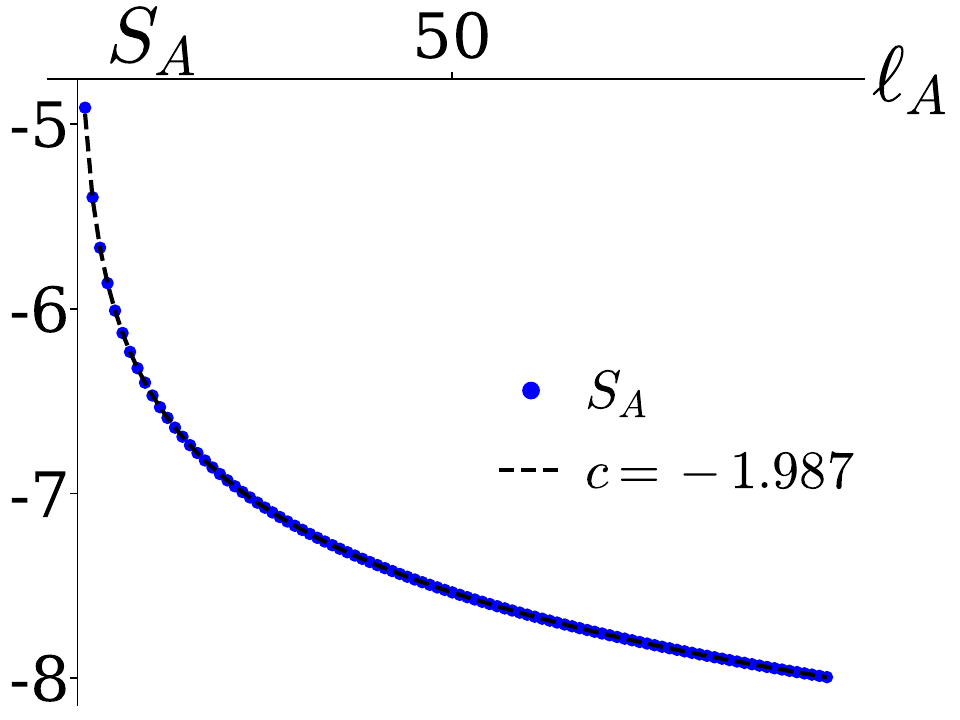}{\linewidth}{(a)}
  \end{minipage}
  \begin{minipage}[b]{0.48\linewidth}
    \panel{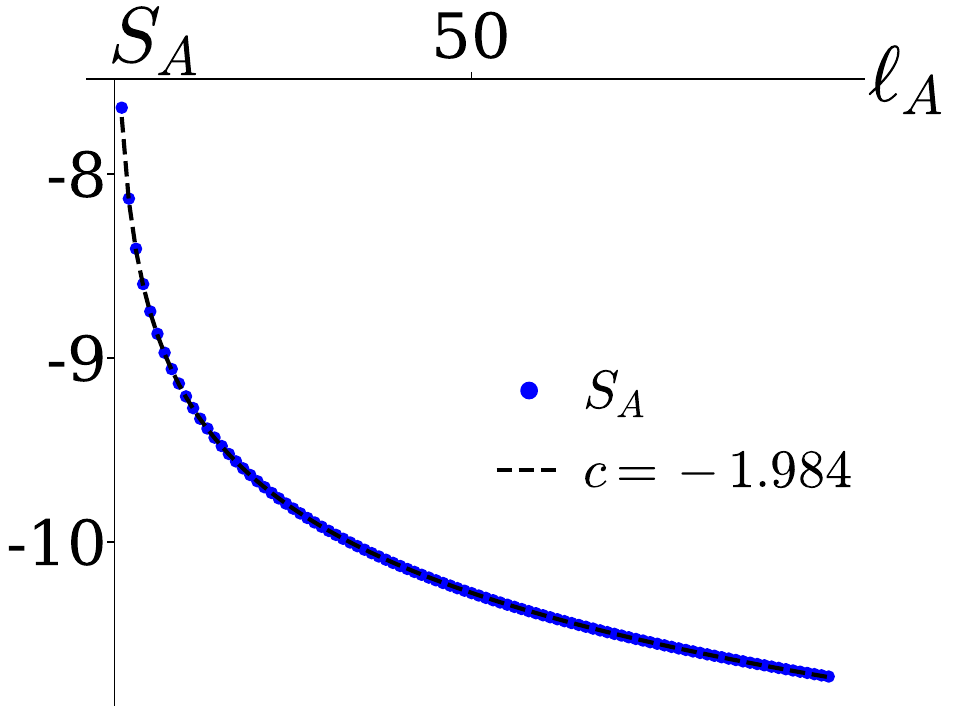}{\linewidth}{(b)}
  \end{minipage}
\caption{
Logarithmic scaling of the interval entropy in two non-Hermitian critical free-fermion models. 
(a) Two-legged non-Hermitian SSH model with \((v,w,u)=(2,1,1-10^{-12})\). 
(b) Two-legged symplectic-fermion model with a small mass detuning \(m=2+10^{-12}\). 
Blue dots denote the numerical data, while the black dashed curves are fits to \(S_A=(c_{\mathrm{fit}}/3)\log\ell_A+s_0\). 
The fitted values \(c_{\mathrm{fit}}\simeq -1.987\) and \(c_{\mathrm{fit}}\simeq -1.984\) are consistent with \(c=-2\).
}
  \label{fig:appEE}
\end{figure}

These two models share the same long-range entanglement structure described by the Cardy form
\begin{equation}
S_A
=
\frac{c}{3}
\log
\left[
\frac{L}{\pi}
\sin\left(\frac{\pi \ell_A}{L}\right)
\right]
+s_0,
\label{eq:app_cft_fit_form_nhssh}
\end{equation}
where $c$ is the central charge, $L$ is the total system size, $\ell_A$ is the subsystem size, and $s_0$ is a nonuniversal constant reflecting UV entanglement. The numerical result shown in Fig.~\ref{fig:appEE}(a) gives the logarithmic coefficient \(c=-2\).

\paragraph{Symplectic-fermion lattice model.}

As a second representative example, we consider the lattice model for symplectic fermions. Following Ref.~\cite{symplectic}, the Hamiltonian is
\begin{equation}
H
=
-\sum_{j=1}^{2L}
i(-1)^j
\left[
c_{j+1}c_j^\dagger
+
c_jc_{j+1}^\dagger
+ 
i
\left(
c_j^\dagger c_j
+
c_{j+1}^\dagger c_{j+1}
\right)
\right].
\label{eq:app_symplectic_real_space}
\end{equation}
Introducing a two-site unit cell, the momentum-space Hamiltonian can be written as
\begin{equation}
\mathcal{H}_{\mathrm{SF}}(k)
=
\begin{pmatrix}
2 & q(k) \\
-q^*(k) & -2
\end{pmatrix},
\quad
q(k)=i(1+e^{ik}).
\label{eq:app_symplectic_bloch}
\end{equation}
The energy dispersion is
\begin{equation}
E_\pm(k)
=
\pm
\sqrt{
2
-
2\cos k
},
\label{eq:app_symplectic_spectrum}
\end{equation}
with the gap closing at \(k=0\). In numerical calculations, a small detuning of the diagonal term, \(m = 2+\delta\) with \(0<\delta \ll 1\), is introduced to regularize the exceptional point. Although this model is not written in a PT-symmetric form, it is pseudo-Hermitian and has a real spectrum for diagonal term \(m\geq 2\).

We also introduce a two-legged version of the symplectic-fermion lattice model, with hopping function 
\begin{equation} 
q(k) = i(1+\cos k). 
\label{eq:app_symplectic_even_factor} 
\end{equation} 
The two-legged version has the same long-range entanglement behavior as the original symplectic-fermion lattice model. Their biorthogonal entanglement entropy follows the Cardy form in Eq.~\eqref{eq:app_cft_fit_form_nhssh}, and the numerical result shown in Fig.~\ref{fig:appEE}(b) again gives \(c=-2\).

The non-Hermitian SSH model and the lattice symplectic-fermion model therefore provide two representative non-Hermitian critical free-fermion systems whose biorthogonal entanglement entropy is governed by the same central charge. The nH SSH model is the boundary lattice model used in the main text, while the symplectic-fermion model provides an additional free-fermion example. In Appendix Sec.~\ref{app:other_free_fermions}, we further show the cMERA computation for the additional example.

\section{Local circuit form of the symmetric left-right information metric}
\label{app:LR_metric}

In the Hermitian cMERA construction, the information metric along the RG direction is defined from the overlap between neighboring states,
\begin{equation}
g_{uu}(u)du^2
=
1-
\left|
\langle \Psi(u)|\Psi(u+du)\rangle
\right|^2 .
\end{equation}
For the non-unitary density-matrix circuit considered in this work, the state at a finite cMERA depth is represented instead by the biorthogonal density matrix
\begin{equation}
\rho(u)
=
|\Psi_R(u)\rangle\langle\Psi_L(u)| .
\end{equation}
Accordingly, in the main text we generalized the overlap metric to the symmetric left-right form
\begin{align}
g_{uu}(u)du^2
=
1-
&\langle \Psi_L(u)|\Psi_R(u+du)\rangle \nonumber\\
\times&\langle \Psi_L(u+du)|\Psi_R(u)\rangle .
\label{eq:app_lr_metric_def}
\end{align}
Since this expression involves the right and left states on the two sides of the density-matrix circuit, its locality along the cMERA direction is not manifest. The purpose of this appendix is to show that, in our case, the generalized definition in Eq.~\eqref{eq:app_lr_metric_def} is equivalent to a local definition controlled only by the entangler \(K(u)\) at the corresponding cMERA layer.

For a normalized biorthogonal pair the neighboring right and left states are related by the infinitesimal cMERA layer,
\begin{align}
|\Psi_R(u+&du)\rangle
\simeq\nonumber\\
&\left[
1-iK(u)du
-\frac{1}{2}K^2(u)du^2
\right]
|\Psi_R(u)\rangle ,
\\
\langle\Psi_L(u+&du)|
\simeq\nonumber\\
&\langle\Psi_L(u)|
\left[
1+iK(u)du
-\frac{1}{2}K^2(u)du^2
\right] .
\end{align}
Here \(K(u)\) denotes the local cMERA entangler in the interaction picture, so the dilation part has already been separated from the entangling part of the flow. We also write
\begin{equation}
\langle O\rangle_{LR}
\equiv
\langle\Psi_L(u)|O|\Psi_R(u)\rangle .
\end{equation}
Expanding Eq.~\eqref{eq:app_lr_metric_def} gives
\begin{equation}
g_{uu}(u)
=
\langle K^2(u)\rangle_{LR}
-
\langle K(u)\rangle_{LR}^2 .
\label{eq:app_metric_variance}
\end{equation}
For the ansatz used in this work, the circuit follows a minimal path: the generator contains no component parallel to the state, a component that would not contribute to disentangling, i.e.,
\begin{equation}
\langle K(u)\rangle_{LR}=0 .
\end{equation}
Therefore the metric obtained from the generalized left-right overlap reduces to
\begin{equation}
g_{uu}(u)
=
\langle K^2(u)\rangle_{LR}.
\label{eq:app_metric_K2}
\end{equation}
In each active momentum sector, the entangler acts on a two-dimensional single-particle space. After removing the identity component, which does not disentangle the state, the minimal local entangler is represented by a traceless \(2\times2\) matrix. For any \(2\times2\) matrix \(K\), the Cayley-Hamilton identity gives
\begin{equation}
K^2
-
(\mathrm{tr}\,K)K
+
(\det K)\mathbf I
=
0 .
\end{equation}
Since the minimal entangler is traceless, we obtain
\begin{equation}
K^2
=
-(\det K)\mathbf I .
\end{equation}
Thus \(K^2\) is proportional to the identity in the two-level sector and is independent of the state on which it acts. The generalized symmetric left-right overlap in Eq.~\eqref{eq:app_lr_metric_def} therefore reduces to the local metric obtained from the entangler data of the circuit, which justifies the local use of the left-right metric in the main text.

\section{Other critical free-fermion examples}
\label{app:other_free_fermions}

In this appendix we provide another critical free-fermion example whose cMERA fixed point gives the same emergent geometry. The model is the two-legged symplectic-fermion representative introduced in Appendix~\ref{app:nH_free_fermions}. Its Bloch Hamiltonian is
\begin{equation}
H_{\mathrm{SF}}^{(2)}(k)
=
\begin{pmatrix}
2 & i(1+\cos k)\\
i(1+\cos k) & -2
\end{pmatrix}.
\end{equation}
The corresponding single-particle spectrum is
\begin{equation}
E_\pm(k)
=
\pm
\sqrt{
4-(1+\cos k)^2
}.
\end{equation}
Thus the gap closes at \(k=0\), where \(1+\cos k=2\).

The occupied right and left eigenstates can be written in the biorthogonal form
\begin{equation}
|R_{k,-}\rangle
=
\begin{bmatrix}
-\sin\frac{\phi_k}{2}\\
\cos\frac{\phi_k}{2}
\end{bmatrix},
\qquad
|L_{k,-}\rangle
=
\begin{bmatrix}
-\sin^*\frac{\phi_k}{2}\\
\cos^*\frac{\phi_k}{2}
\end{bmatrix},
\end{equation}
where the complex band angle is
\begin{equation}
\phi_k
=
\arctan\left(
\frac{i(1+\cos k)}{2}
\right).
\end{equation}

We now specialize the fermionic cMERA construction to the two-legged symplectic-fermion representative. We adopt the same single-angle cMERA ansatz,
\begin{equation}
M(k,u)
=
\begin{pmatrix}
\cos\theta_k(u) & \sin\theta_k(u)\\
-\sin\theta_k(u) & \cos\theta_k(u)
\end{pmatrix}.
\end{equation}
The boundary conditions are fixed by matching the UV state at \(u=0\) and requiring the circuit to flow to a simple product reference state as \(u\to-\infty\). Concretely,
\begin{equation}
\theta_k(-\infty)
=
\frac{\phi_{k=\pi}}{2} = 0,
\qquad
\theta_k(0)
=
\frac{\phi_k}{2},
\end{equation}

Since the rotation angle is even in momentum, the corresponding cMERA disentangler is chosen in the even lattice channel. We take
\begin{equation}
iK(k,u)
=
\begin{pmatrix}
0 & -g(u)\\
g(u) & 0
\end{pmatrix}
\Gamma\left(\frac{|k|e^{-u}}{\pi}\right),
\end{equation}
where \(\Gamma\) is the same running cutoff used in the main text. The accumulated rotation of a mode \(k\) is fixed by
\begin{equation}
\frac{\phi_k}{2}
=
\int_0^{\ln(k/\pi)}
g(s)\,ds .
\end{equation}
Differentiating this relation with respect to \(k\), and using \(k=\pi e^u\), gives
\begin{equation}
g(u)
=
-\frac{k}{2}
\partial_k \phi_k
\bigg|_{k=\pi e^u}.
\end{equation}

The long-wavelength form of the phase is
\begin{align}
\phi_k
\simeq
-i\log k
+
\frac{i}{2}\log 8
+
O(k^2).
\end{align}
Substituting this expansion into the layer kernel gives a constant kernel in the infrared
\begin{equation}
g(u) = \frac{i}{2}.
\end{equation}
The sign depends on the orientation convention for the local rotation. The metric component is therefore
\begin{equation}
g_{uu}
=
g^2(u)
=
-\frac{1}{4}.
\end{equation}
Thus the two-legged symplectic-fermion representative produces the same de Sitter cMERA fixed point as the two-legged non-Hermitian SSH model. This provides another non-Hermitian critical free-fermion example whose long-wavelength entanglement-renormalization geometry is \(dS_2\).

The same long-wavelength conclusion also applies to the original lattice conventions reviewed in Appendix Sec.~\ref{app:nH_free_fermions}. For the original non-Hermitian SSH model and the original lattice symplectic-fermion model, $q_{\mathrm{SSH}}(k)=v-we^{-ik}$, $q_{\mathrm{SF}}(k)=i(1+e^{ik})$,
the occupied state is characterized not only by the non-Hermitian angle associated with \(|q(k)|\), but also by the momentum-dependent phase \(\arg q(k)\). The corresponding cMERA rotation therefore generally involves more than one Pauli-matrix direction, unlike the single-channel two-legged representatives used for the explicit computations above. This difference belongs to the UV lattice representation. Near the critical momentum, the additional phase is analytic, while the exceptional-point part of the band angle has the same logarithmic singularity as in the two-legged models. Hence the original models share the same long-wavelength cMERA kernel and the same deep-bulk \(dS_2\) metric. Together with the two-legged symplectic-fermion computation, these examples show that the connection between non-Hermitian critical free fermions and emergent de Sitter geometry extends beyond the specific lattice model used in the main text.

\section{Massive check}
\label{app:massive}

In this appendix we provide an additional check away from criticality by considering a massive deformation of the two-legged non-Hermitian SSH model.
For parameters away from the critical point, the spectrum has a finite gap \(\Delta\). In this case the entanglement entropy no longer obeys
critical logarithmic scaling with the interval length, and the interval dependence is more sensitive to short-distance details. We therefore focus on
the saturation value of the entanglement entropy.

\begin{figure}
    \includegraphics[width=0.8\linewidth]{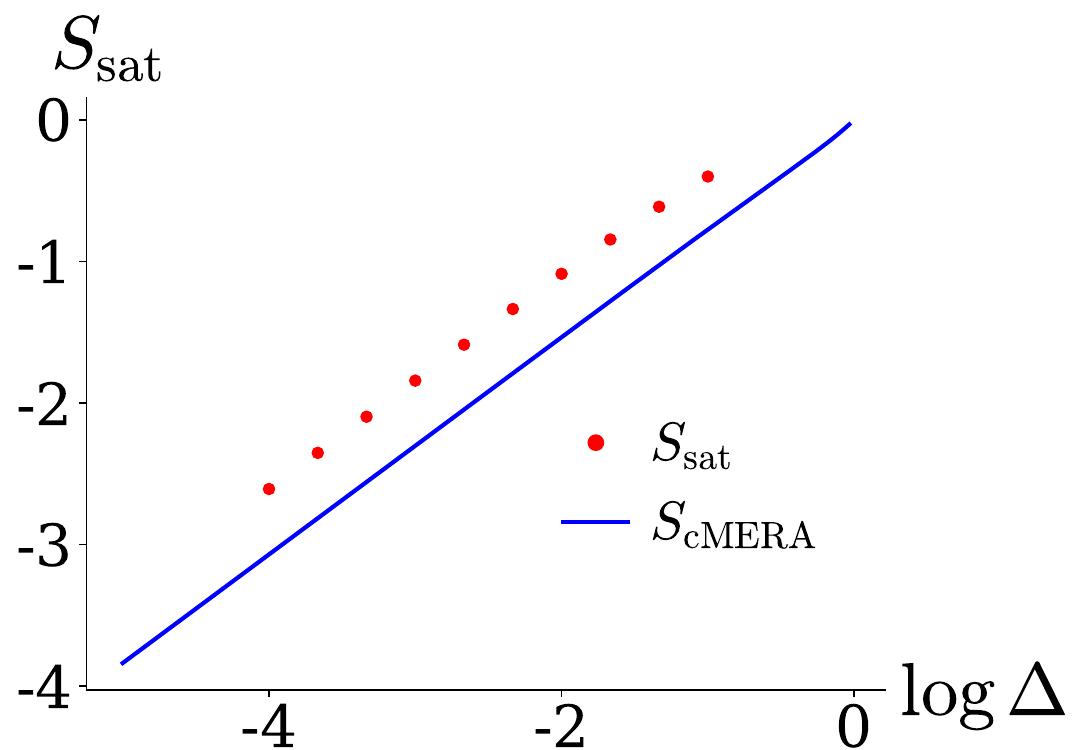}
    \caption{
    Saturation entropy \(S_{\mathrm{sat}}\) of the massive non-Hermitian two-legged SSH model as a function of the energy gap \(\Delta\), with \((v,w)=(2,1)\).
    The red dots are lattice results, and the blue line is the vertical-geodesic cMERA prediction \(S_{\mathrm{cMERA}}\), normalized by the critical-case factor \(\lambda=2/3i\).
    The agreement up to an additive constant provides a gapped-case consistency check of the entropy/geodesic matching.
    }
    \label{fig:massive}
\end{figure}

The overall normalization between the entanglement entropy and the cMERA geodesic length is fixed from the critical comparison. Using the same ratio, \(\lambda=2/3i\), we compare the massive cMERA/geodesic result with the exact saturation entropy. As shown in Fig.~\ref{fig:massive}, the two curves agree as functions of the gap scale up to an additive constant. This provides a gapped-case consistency check of the extremal-length prescription.

\bibliography{reference}

\end{document}